\documentclass{aa}  

\usepackage{graphicx}
\usepackage[dvipsnames]{xcolor}
\usepackage{txfonts}
\usepackage{mathrsfs}
\usepackage[normalem]{ulem}
\usepackage{longtable}
\usepackage{caption}
\usepackage{geometry}
\usepackage{booktabs}
\usepackage{lscape}
\usepackage{booktabs}

\newcommand{\aFe}{[\alpha/\mathrm{Fe}]}
\newcommand{\Mo}{M_\sun}
\newcommand{\MMS}{M_{\mathrm{MS}}}
\newcommand{\Mmin}{M_\mathrm{min}}
\newcommand{\MHB}{M_\mathrm{HB}}
\newcommand{\Teff}{T_\mathrm{eff}}
\newcommand{\McHe}{M_\mathrm{cHe}}
\newcommand{\DYRGB}{\Delta Y_{\text{RGB}}}
\newcommand{\lLRR}{\log L^{\text{RRL}}_{3.83}}
\newcommand{\MRR}{M^{\text{RRL}}_{3.83}}
\newcommand{\FeH}{[\mathrm{Fe}/\mathrm{H}]}

\begin{document} 

\title{The PGPUC horizontal branch evolutionary tracks}

   \author{Aldo A. R. Valcarce\inst{1}
   \and
    Márcio Catelan\inst{2,3,4}
    \and
    Sânzia Alves\inst{1} 
    \and
    Felipe González-Bordon\inst{1} 
         }

   \institute{
            Departamento de Física, FACI, Universidad de Tarapacá, Casilla 7D, Arica, Chile
            \and
            Instituto de Astrofísica, Pontificia Universidad Católica de Chile, Av. Vicuña Mackenna 4860, 7820436 Macul, Santiago, Chile 
            \and
            Millennium Institute of Astrophysics, Nuncio Monseñor Sotero Sanz 100, Of. 104, Providencia, Santiago, Chile
            \and
            Centro de Astro-Ingeniería, Pontificia Universidad Católica de Chile, Av. Vicuña Mackenna 4860, 7820436 Macul, Santiago, Chile
             }

   \date{Received XXXX XX, 2025; accepted XXXX XX, 202X}

 
  \abstract
   {The horizontal branch (HB) phase of stellar evolution plays a critical role in understanding the life cycle of stars, particularly in the context of low-mass stars. However, generating theoretical evolutionary tracks for HB stars is computationally intensive, especially when attempting to match a wide range of observational data.}   
   {This work presents an extensive grid of HB evolutionary tracks computed using the PGPUC code, covering a broad range of chemical compositions, progenitor masses, and alpha-element distributions. The aim is to provide a robust tool for interpreting HB stellar populations and advancing our understanding of their diverse properties. The evolutionary tracks are made publicly available through the PGPUC Online database for easy access and interpolation. }
   {We computed over 19\,000 HB evolutionary tracks encompassing a wide range in terms of mass, metallicity, helium abundance, and alpha-element enhancement, along with different progenitor masses. Using these tracks, we calculated zero-age, middle-age, and terminal-age loci.}
   {The PGPUC database now includes a fine grid of HB evolutionary tracks, allowing for precise interpolation. Key findings include the dependence of HB morphology on progenitor mass, metallicity, and helium abundance. 
   }
   {}

   \maketitle

\section{Introduction}
\label{sec:introduction}

The study of stellar evolution, particularly the horizontal branch (HB) phase, is crucial for understanding the life cycle of stars. The HB phase, characterized by helium core burning in low- to intermediate-mass stars, follows the red giant branch (RGB) phase and precedes the asymptotic giant branch (AGB) phase. This evolutionary stage is marked by a diverse range of stellar properties, influenced by initial chemical compositions and evolutionary histories, which makes it a key focus in astrophysical research \citep[e.g.,][]{Salaris_Cassisi2005, Catelan2009}.

In this sense, stellar evolutionary tracks are essential tools for interpreting observational data and understanding stellar populations. However, creating these tracks is computationally intensive and time-consuming, especially when a wide spectrum of initial parameters must be considered to match the observed diversity in stellar populations. Historically, researchers have dedicated significant resources to developing comprehensive sets of evolutionary tracks \citep[e.g.,][]{Pietrinferni_etal2004, VandenBerg_etal2006, Dotter_etal2008, Valcarce_etal2012}. These tracks typically vary in mass, global metallicity ($Z$), initial helium abundance ($Y$), and alpha-element enhancement ($\aFe$), reflecting the different chemical compositions observed in stars.

The progenitor mass ($\MMS$), defined here as the mass of the main sequence (MS) star that has just reached the RGB tip at a given age, is another critical parameter influencing the HB phase, particularly affecting the helium core mass ($\McHe$). Understanding the impact of progenitor mass variations on HB stars is essential to accurately model their evolutionary paths and properties \citep{Rood1973}. Despite the availability of various sets of evolutionary tracks, the need for a rapidly accessible and comprehensive database of HB evolutionary tracks remains pressing.

In this paper, we present a new extensive set of HB evolutionary tracks computed using the Princeton-Goddard-Pontificia Universidad Católica stellar evolution code \citep[PGPUC SEC,][]{Valcarce_etal2012}. This dataset includes over $19\,000$ evolutionary tracks with varying chemical compositions and progenitor masses, providing a robust tool for studying the effects of these parameters on HB stars. We have computed zero-age, middle-age, and terminal-age HB loci to investigate the impact of changing initial conditions on HB stellar populations. These evolutionary tracks are integrated into the PGPUC Online database, designed for fast interpolation and downloading, thereby facilitating efficient comparison of theoretical models with observational data.

This paper details the methodology used to generate the evolutionary tracks, the range of parameters considered, and a potential application. It is organized as follows. First, in Sect. \ref{sec:PGPUCsec} we describe the code used for creating the evolutionary tracks, followed by Sect. \ref{sec:HBmodels}, which presents the HB evolutionary tracks created. Then, in Sect. \ref{sec:TimeLoci} we describe the three main loci~--- zero-age, middle-age-, and terminal-age~--- that were created with these HB models, followed by the conclusions in Sect. \ref{sec:conclusions}.

\section{The PGPUC stellar evolution code}
\label{sec:PGPUCsec}
The PGPUC SEC, described in full in \citet{Valcarce_etal2012}, represents an updated version of the code created by M. Schwarzschild and R. Härm at Princeton University \citep{Schwarzschild_Harm1965, Harm_Schwarzschild1966}, and then extensively modified by Allen V. Sweigart at NASA's Goddard Space Flight Center \citep[][and references therein]{Sweigart1971, Sweigart1973, Sweigart1994, Sweigart1997, Sweigart_Demarque1972, Sweigart_Gross1974, Sweigart_Gross1976, Sweigart_Catelan1998}. The PGPUC SEC serves as a robust tool for primarily investigating low-mass stars spanning from the MS to the white dwarf cooling sequence. Low-mass stars are defined as those that develop a degenerate helium core following the MS evolution, corresponding to stars with masses below 2 or 3 $\Mo$, depending on their chemical composition.

\subsection{Input physics}

The physical inputs utilized in the PGPUC code include:
\begin{itemize}
\item Opacity for high temperatures in a tabular format from \citet{Iglesias_Rogers1996} and for low temperatures from \citet{Ferguson_etal2005};
\item Conductive opacities from \citet{Cassisi_etal2007}; 
\item Neutrino energy losses based on the fitting formula by \citet{Haft_etal1994};
\item Thermonuclear reaction rates derived from the NACRE compilation \citep{Angulo_etal1999}, with the exception of the $^{12}{\rm C}(\alpha,\gamma)^{16}{\rm O}$ reaction, which is sourced from \citet{Kunz_etal2002};
\item \citeauthor{Irwin2007}'s (\citeyear{Irwin2007}) FreeEOS equation of state (EOS), option EOS4. The impact of this EOS on HB stellar models is discussed in \citet{Cassisi_etal2003}; 
\item Boundary conditions derived from \citet{Krishna_Swamy1966}, fitted to a refined formula presented in \citet{Valcarce_etal2012};
\item Mass loss formula adapted from \citet{Schroder_Cuntz2005} for the RGB phase (mass loss is turned off during the HB phase);
\item Electron screening of nuclear reactions from \citet{Dewitt1973} and  \citet{Graboske_etal1973}.\footnote{Unlike what has been stated in some recent studies, PGPUC does not use electron screening from \citet{Salpeter1954}. Further details regarding this issue can be found in Appendix~\ref{sec:appendix}.}
\end{itemize}

Proper treatment of semiconvection at the outer edge of the convective core is crucial to represent the HB phase correctly \citep{Demarque_Mengel1972}. The PGPUC SEC treats this semiconvection with the procedure described in \citet{Sweigart_Demarque1972}. The treatment for suppressing the so-called breathing pulses is implemented as described in \citet{Sweigart1971}.

Horizontal branch models include an amount of extra helium in the envelope, typically $\DYRGB \leq 0.02$, depending on $Z$, $Y$, and $\MMS$ (see Table~\ref{TableC}). This enrichment is associated with the occurrence of the first dredge-up during the RGB stage, which brings helium-rich material from the interior to the outer layers of the star \citep[e.g.,][]{Iben1964,Renzini1977,Sweigart_Gross1978}.

Updates to some of these input physics ingredients have been provided in the literature in recent years. If taken into account, they would introduce small modifications to the resulting theoretical models and evolutionary tracks. In this work, we choose to maintain the original input physics \citep[as in][]{Valcarce_etal2012} to ensure consistency with the existing PGPUC Online database. By keeping the same physical assumptions, we preserve the homogeneity between previously published PGPUC models and the new ones presented in this paper, ensuring that no discrepancies arise from differences in the input physics.

\begin{figure*}
    \centering
        \includegraphics[width=2.0\columnwidth]{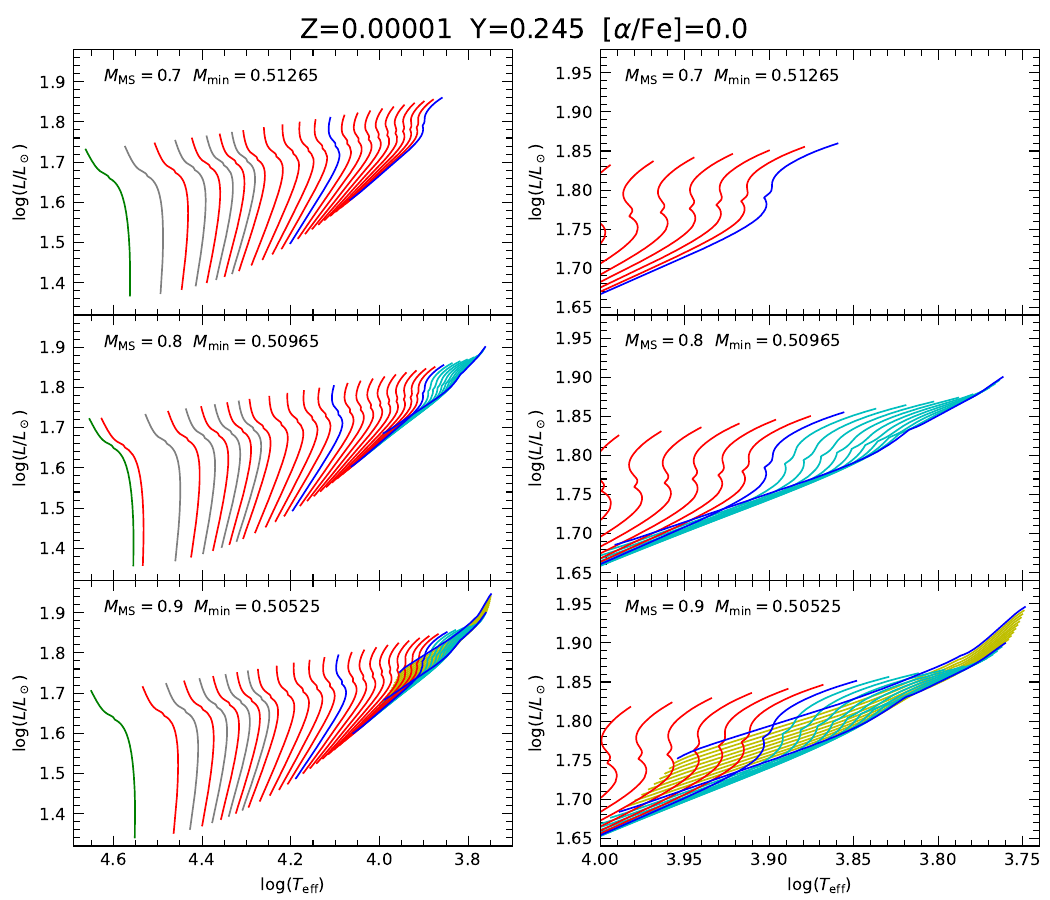}
    \caption{HB evolutionary tracks in the theoretical plane~--- luminosity, $\log(L/L_\sun)$, vs. effective temperature, $\log \Teff$~--- for $Z=0.00001$, $Y=0.245$, and $\aFe=0.00$. The top, middle, and bottom left panels show the set of evolutionary tracks corresponding to progenitor masses of $\MMS=0.700\,\Mo$, $0.800\,\Mo$, and $0.900\,\Mo$, respectively. The right panels are a zoom-in around the cooler part of the HB. Each panel also shows the minimum mass, $\Mmin$, reached for each set (green lines), corresponding to the hottest track. Blue, red, and gray lines correspond to evolutionary tracks in steps of $\Delta M=0.100\,\Mo$, $\Delta M=0.010\,\Mo$ (for masses lower than $0.700\,\Mo$), and $\Delta M=0.005\,\Mo$ (for masses lower than $0.550 \Mo$), respectively. Cyan lines correspond to evolutionary tracks computed with masses greater than $0.700 \, \Mo$ and lower than $0.800\,\Mo$, in steps of $\Delta M=0.010\,\Mo$. Yellow lines correspond to evolutionary tracks for masses greater than $0.800 \, \Mo$ and lower than $0.900\,\Mo$, in steps of $\Delta M=0.010\,\Mo$.}
    \label{Fig:HB1}
\end{figure*}
\begin{figure*}
    \centering
        \includegraphics[width=2.0\columnwidth]{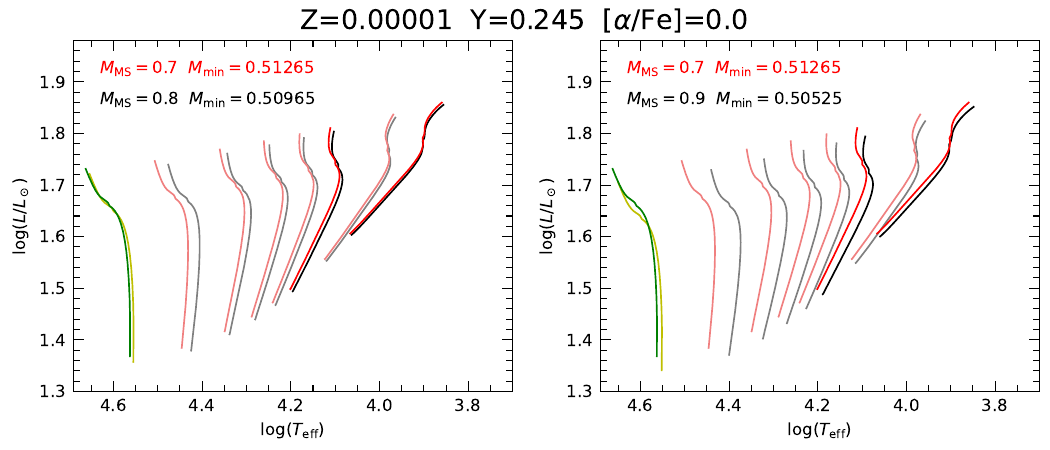}
    \caption{Comparison between HB evolutionary tracks for $Z=0.000010$, $Y=0.245$, and $\aFe=0.00$ with $\MMS=0.700\,\Mo$ and: i) $\MMS=0.800\,\Mo$ (left panel), and ii) $\MMS=0.900\,\Mo$ (right panel). In each panel, green and yellow lines correspond to the $\Mmin$ value reached for $\MMS=0.700 \, \Mo$ and the alternative $\MMS$ value shown in the plot, respectively. Red and black lines correspond to masses in steps of $\Delta M=0.010 \Mo$ for $\MMS=0.700\,\Mo$ and the alternative $\MMS$ values, respectively. In like vein, red and gray lines correspond to evolutionary tracks for $\MMS=0.700\,\Mo$ and the alternative $\MMS$ values, respectively, with masses of $\MHB=0.650\,\Mo$, $0.580\,\Mo$, $0.560\,\Mo$, $0.540\,\Mo$, $0.520\,\Mo$, and in steps of $\Delta M=0.010\,\Mo$ for sets with evolutionary tracks that reach masses lower than $0.500 \, \Mo$.
   }
    \label{Fig:HB1b}
\end{figure*}

\subsection{The PGPUC Online database}

The PGPUC Online database\footnote{\url{http://www2.astro.puc.cl/pgpuc/}} enables users to interpolate evolutionary tracks from the MS to the tip of the RGB, isochrones, and zero-age horizontal branch (ZAHB) loci from a finely sampled grid of theoretical evolutionary tracks created utilizing the PGPUC SEC. Initially, this grid encompassed masses from $M=0.5\,\Mo$ to $M=1.1\,\Mo$, initial helium abundances from $Y=0.230$ to $Y=0.370$, and global metallicities from $Z=0.00016$ to $Z=0.01570$, with a fixed distribution of alpha-elements of $\aFe= \log(\alpha/{\rm Fe}) -\log(\alpha/{\rm Fe})_{\odot} = 0.3$ for a \citet{Grevesse_Sauval1998} solar chemical composition, with our solar models being characterized by $(Z, \, Y)_{\odot} = (0.0167, \, 0.262)$ \citep{Valcarce_etal2012}. Mass loss follows the \citet{Schroder_Cuntz2005} prescription, with $\eta = 1.0$. Further details can be found in \citet{Valcarce_etal2012}. Over the past decade, this grid has been expanded to include models for $\aFe=0.0$, and a mass loss rate with $\eta = 0.0$, as well as extending metallicities from $Z=0.00001$ to $Z=0.06000$ for evolutionary tracks and isochrones, and from $Z=0.00001$ to $Z=0.03000$ for ZAHB loci.

An essential input parameter for creating a ZAHB locus is $\MMS$, which, together with $Z$, $Y$, and $\aFe$, determines the $\McHe$ value at the tip of the RGB, and thus at the ZAHB. The $\MMS$ parameter is particularly relevant for accurately predicting the position and shape of the ZAHB at high $\Teff$ (e.g., $\Teff>11\,500$~K) and defining the evolution throughout the HB phase \citep{Eggleton1968, Sweigart_Gross1976, Sweigart_Gross1978, Vandenberg1992}. Higher values of this parameter ($\MMS>1.0\,\Mo$) also influence the location of HB models at the red end. For a fixed chemical composition, HB models originating from higher $\MMS$ progenitors tend to appear redder and more luminous than their lower-mass counterparts. This shift has important consequences for their pulsation properties \citep[e.g.,][]{Cassisi_etal1997}.

The models provided by this database are available in the theoretical Hertzsprung–Russell diagram as well as in various photometric filters that can be chosen by the user. Currently, users have a choice from about 50 filter systems, primarily derived from bolometric corrections by \citet{Chen_etal2019}, as well as by \citet{Casagrande_VandenBerg2018}.

Moreover, this database includes two calculators for determining metallicity values and iron-over-hydrogen ratios ([Fe/H]). One calculator enables users to obtain the metallicity, $Z$, from a given [Fe/H] value, helium mass fraction, $Y$, and alpha-element enhancement ratio, $\aFe$,\footnote{\url{http://www2.astro.puc.cl/pgpuc/Zcalculator.php}} while the other allows for the reverse calculation for obtaining [Fe/H] from $Z$, $Y$, and $\aFe$.\footnote{\url{http://www2.astro.puc.cl/pgpuc/FeHcalculator.php}} In the following section, we present a significant expansion of this grid to encompass the evolution of HB stars from the ZAHB to the terminal-age HB (TAHB).

\section{Description of HB tracks}
\label{sec:HBmodels}
The HB evolutionary tracks (hereafter, the HB tracks) were calculated using the same grid of chemical compositions as the ones employed in the set of evolutionary tracks spanning from the MS to the tip of the RGB available in the PGPUC Online database. 
Specifically, this entails:

\begin{itemize}
    \item Metallicity by mass: $Z=0.00001$, $0.00016$, $0.00028$, $0.00051$, $0.00093$, $0.00160$, $0.00284$, $0.00503$, $0.00890$, $0.01570$, and $0.03000$.
    \item Helium abundance by mass: $Y=0.230$, $0.245$, $0.270$, $0.295$, $0.320$, $0.345$, and $0.370$.
    \item Alpha-element enhancement ratio: $\aFe=0.0$ and $0.3$.
\end{itemize}

\begin{figure}
    \centering
        \includegraphics[width=\columnwidth]{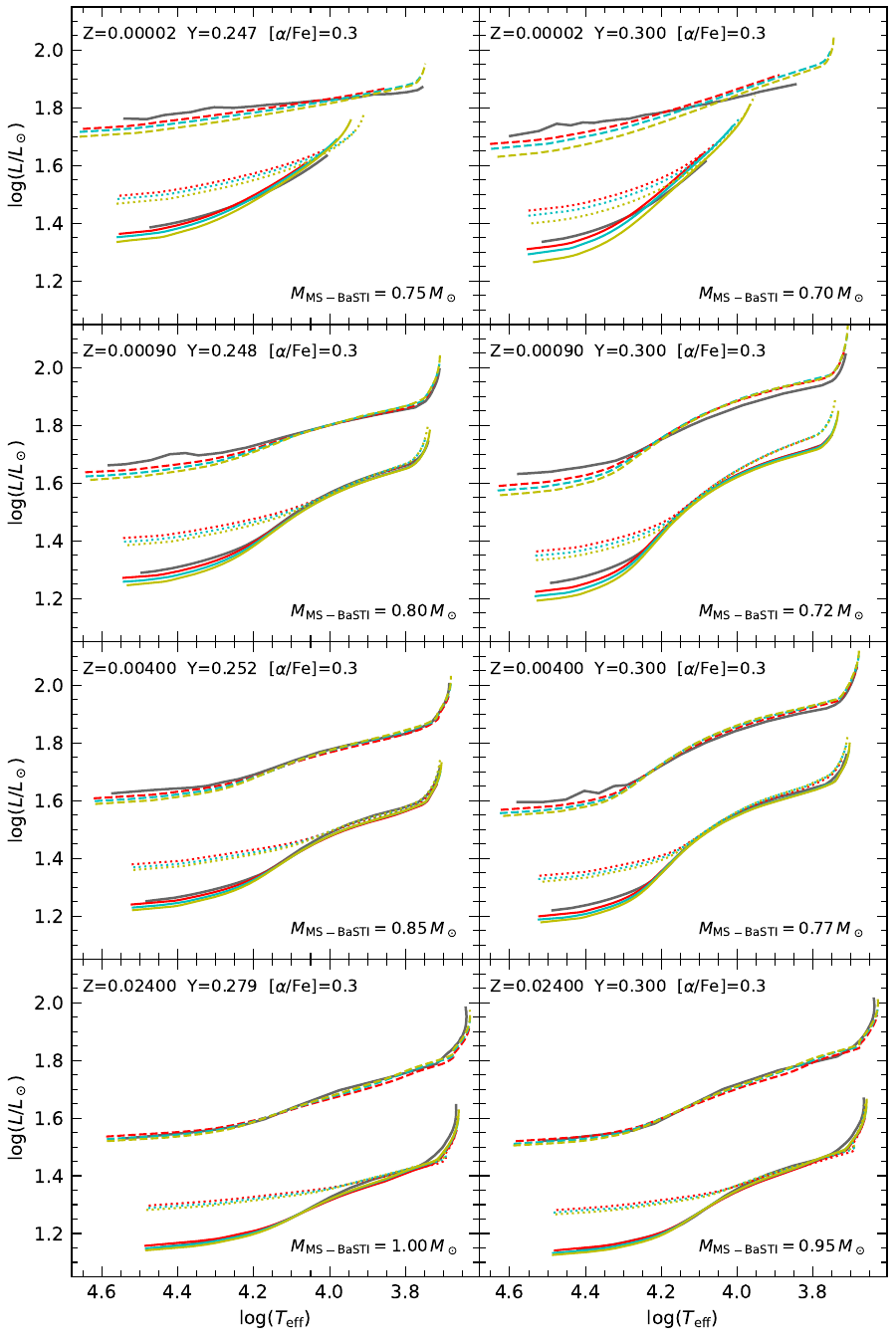}
    \caption{PGPUC ZAHB (continuous lines), MAHB (dotted lines), and TAHB loci (dashed lines) in the theoretical plane for eight chemical compositions listed in each panel. Red, cyan, and yellow lines correspond to progenitor masses of $0.700$, $0.800$, and $0.900\, \Mo$, respectively. In gray are shown ZAHB and TAHB loci from BaSTI for comparison, with the progenitor mass of those models indicated in each panel. 
    }
    \label{Fig:PAHBloci}
\end{figure}

\begin{figure}
    \centering
        \includegraphics[width=\columnwidth]{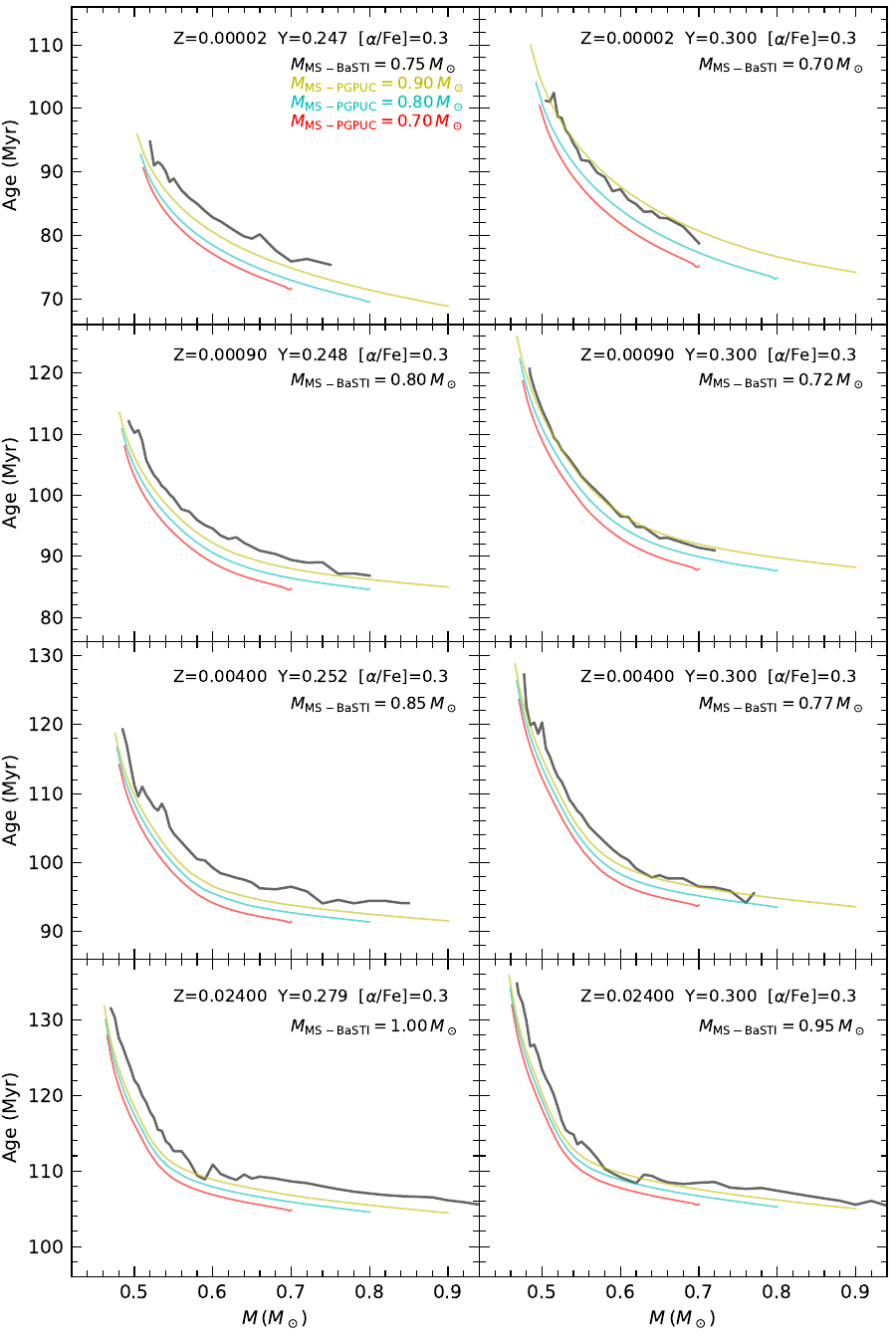}    
    \caption{TAHB age values for HB stars  
    as a function of the HB mass. Red, cyan, and yellow lines correspond to progenitor masses of $0.700$, $0.800$, and $0.900\, \Mo$, respectively, while in gray are shown BaSTI TAHB loci for comparison, with the progenitor mass of the BaSTI models indicated in each panel.  
    }
    \label{Fig:PAHBmassAge}
\end{figure}

\begin{figure}
    \centering
    \includegraphics[width=0.92\columnwidth]{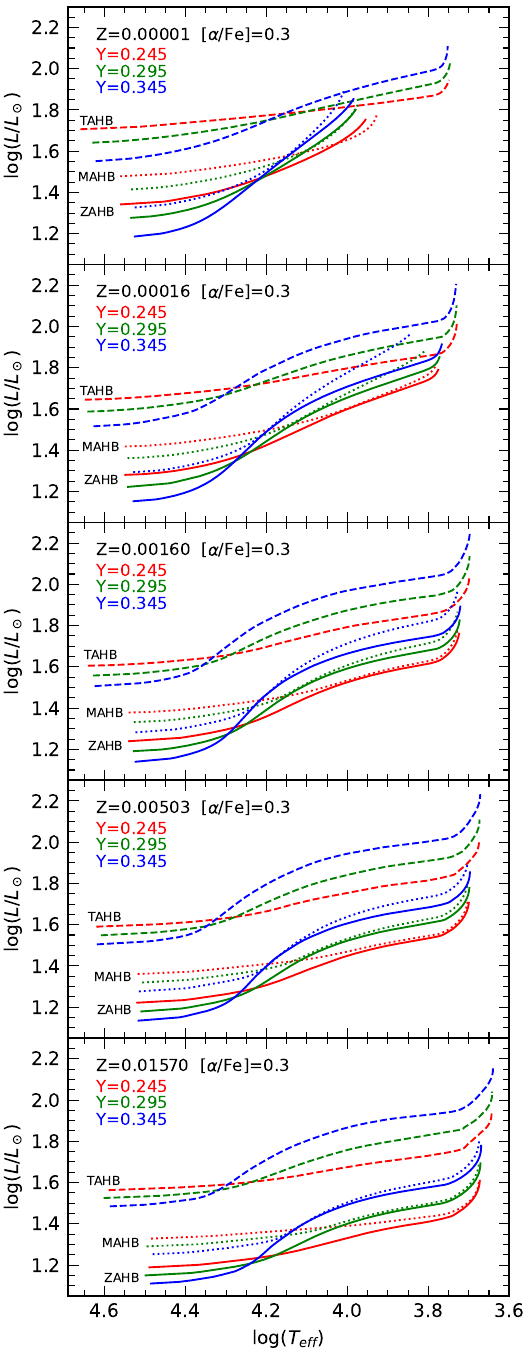}
    \caption{ZAHB (continuous lines), MAHB (dotted lines), and TAHB loci (dashed lines) in the theoretical plane for five metallicities, increasing from the top to the bottom panels, for $\MMS=0.900\, \Mo$ and $\aFe=0.3$. Red, green, and blue lines correspond to helium abundances of $Y=0.245$, $0.295$, and $0.345$, respectively. }
    \label{Fig:Yloci}
\end{figure}

\begin{figure}
    \centering
        \includegraphics[width=0.92\columnwidth]{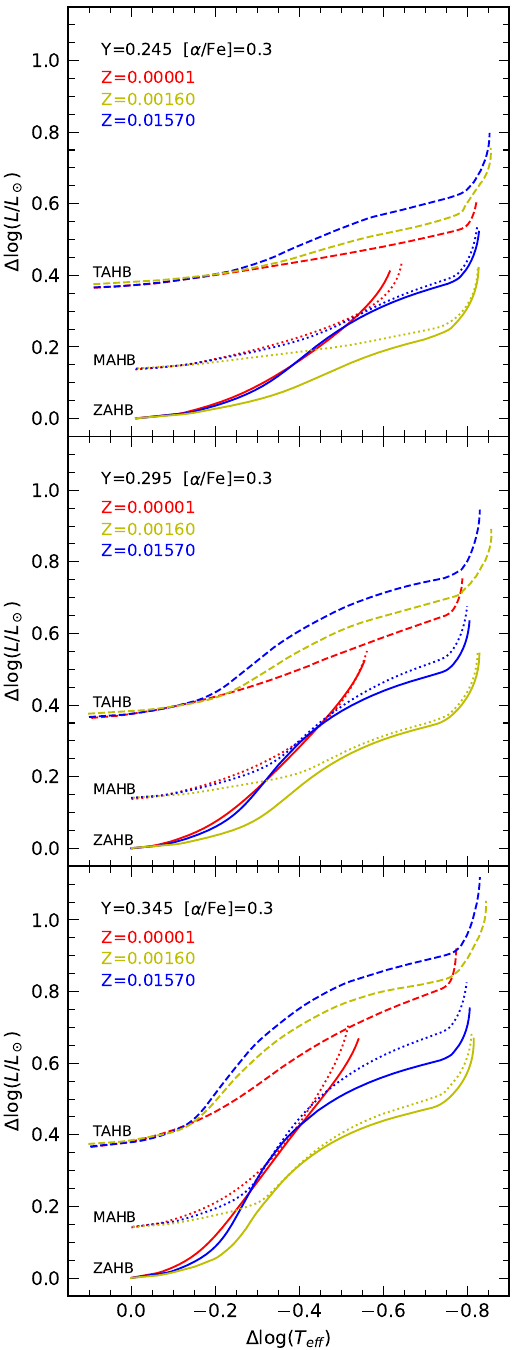}
    \caption{Comparison of ZAHB (continuous lines), MAHB (dotted lines), and TAHB loci (dashed lines) in the theoretical plane for three metallicities ($Z=0.00001$, $0.00160$ and $0.01570$ in red, blue, and yellow, respectively) and three helium abundances ($Y=0.245$, $0.295$, and $0.345$ in the top, middle, and bottom panels, respectively). In all displayed cases,  $\MMS=0.900\,\Mo$ and $\aFe=0.3$ have been assumed. All models have been shifted in $\log(L/L_\sun)$ and $\log \Teff$ such that the highest $\Teff$ of each ZAHB locus aligns at the origin. }
    \label{Fig:Zloci}
\end{figure}

We also included three different MS progenitor masses, namely $\MMS=0.900$, $0.800$, and $0.700$ $\Mo$, for each set of chemical compositions. This was done to account for the effect of age on a given stellar population, which is reflected in the initial $\McHe$ shared by all HB tracks for a given chemical composition and age.  
In this work, the evolution of the progenitor was computed from the MS to the ZAHB without mass loss. Thus, the maximum mass of a set of HB tracks is equal to the corresponding progenitor mass.

The masses used for the computations in a set of HB tracks (hereafter, $\MHB$) range from the mass of the progenitor down to $0.550 \, \Mo$ in steps of $\Delta M=0.010 \, \Mo$, and from $0.550\, \Mo$ down to the lowest mass in each set ($\Mmin$, which differs from $\McHe$ by $\lesssim 0.0002\, \Mo $, in the sense that $\McHe$ is slightly smaller), in steps of $\Delta M=0.005 \, \Mo$. Note that the $\Mmin$ value is different for different chemical compositions and $\MMS$ values.

With all these variables considered, we computed a total of $19\,441$ HB tracks using the Geryon~2 supercomputer\footnote{Geryon 2 is located at the Instituto de Astrofísica and managed in collaboration with the Centro de Astro-Ingeniería at Pontificia Universidad Católica de Chile.}, which amounted to a total of $1\,620$ days of CPU time. All these tracks are included in the PGPUC Online database, providing a smooth grid for interpolating any parameter within the grid.

Figure~\ref{Fig:HB1} shows a set of computed HB tracks for $Z=0.00001$, $\aFe=0.0$, and $Y=0.245$ and three $\MMS$ values, demonstrating the smoothness of our grid. Figure~\ref{Fig:HB1b} shows the comparison between HB tracks with $\MMS=0.800$ and $0.900 \, \Mo$ against $0.700 \, \Mo$ for the same chemical composition as in Fig.~\ref{Fig:HB1}. 

The HB track with $\Mmin$ is shown in each case; the models comprise a small H-rich envelope layer sitting on top of a core of initial mass $\McHe$.  
The behavior of the HB tracks, including the loops observed at high metallicities and high helium abundances, closely aligns with that observed in other studies \citep[e.g.,][]{Sweigart_Gross1976, Bono_etal1997a, Bono_etal1997b, Dotter_etal2007, Pietrinferni_etal2013}. Similar figures, but for other chemical compositions, can be found in Appendix~\ref{sec:extrafigures}.

Figure~\ref{Fig:HB1b} reveals, as was expected, that the differences between the tracks computed for a fixed $\MHB$ but different $\MMS$ values increases as the difference in $\MMS$ increases. This is particularly evident at lower $\MHB$ values. This is due to the fact that, for larger $\MMS$ values, the evolution progresses more rapidly along the RGB phase, resulting in a smaller $\McHe$ values \citep[e.g.,][]{Sweigart_Gross1978}.

As was previously discussed by other authors \citep[][among others]{Faulkner1966, Iben_Rood1970, Gross1973, Sweigart_Gross1978, Dorman1992}, the $\McHe$ value at the tip of the RGB is a physically well-defined quantity that depends sensitively on the progenitor's initial chemical composition and mass, as well as the adopted input physics. While earlier studies have typically reported $\McHe$ values for a limited set of input parameters, our updated PGPUC database provides this information systematically for a broad grid of compositions and $\MMS$. Table~\ref{TableC} lists $\McHe$, the age at the RGB tip, along with the associated helium enrichment in the envelope and both the ZAHB luminosity and mass at $\log \Teff=3.83$ (a characteristic temperature of the RR Lyrae instability strip) for each progenitor mass, $\MMS$, and chemical composition. These data show that $\McHe$ generally decreases with increasing $Y$ and $\MMS$, with a lower variation when $Z$ increases. This structured grid allows for a more continuous and accurate mapping of core masses across different stellar populations, and serves as a robust basis for the construction of HB models.

In order to facilitate the application of PGPUC models in stellar population studies, single HB tracks or complete sets of HB tracks covering the full mass range, from the selected $\MMS$ to $\Mmin$, are available for bulk download from the HB section of the PGPUC Online webpage\footnote{\url{http://www2.astro.puc.cl/pgpuc/hb.php}}. In order to avoid creating large files and make it easier to download the sets of HB tracks, the mass steps, $\Delta \MHB$, were chosen as follows: $\Delta \MHB = 0.0100\,\Mo$ down to $\MHB = 0.550\,\Mo$, after which it decreases to $\Delta \MHB = 0.0050\,\Mo$ for lower masses. Sets of HB tracks with smaller mass steps can be requested via private communication.

Using these HB tracks, it is also possible to calculate the ZAHB, middle-age horizontal branch (MAHB), and TAHB loci for any given chemical composition, which can also be downloaded from the PGPUC Online database. These loci are described in detail in the following section.

\section{ZAHB, MAHB, and TAHB loci}
\label{sec:TimeLoci}
To facilitate comparisons between observations and theory, we have used HB tracks to define, as we have done in previous studies \citep{Catelan_etal2009, Valcarce_etal2014, Valcarce_etal2016}, three key reference loci~--~ZAHB, MAHB, and TAHB~-- each corresponding to a different evolutionary moment in the HB phase. In particular, in addition to the ZAHB, we used two main percentage time loci, following their original definitions given in \citet{Catelan_etal2009}:

\begin{itemize}
    \item TAHB loci, corresponding, for a given chemical composition and $\MMS$ value, to the geometrical locus connecting HB stars with different ZAHB masses in the luminosity (absolute magnitude) versus effective temperature (color) plane when the He at their cores is depleted;
    \item MAHB loci, which are analogous to the TAHB loci, except that they are computed at the halfway point (in terms of age) of the entire HB evolutionary phase.  
\end{itemize}

It is important to note that the MAHB loci correspond to different absolute ages for each HB mass at any given chemical composition, since HB stars with lower masses evolve more slowly than their more massive counterparts \citep{Sweigart_Gross1976,Castellani_etal1994,Zoccali_etal2000}. Examples of these loci are presented in Fig.~\ref{Fig:PAHBloci} within the theoretical ($\log L, \log \Teff$) plane, showing a diverse set of chemical compositions. The dependence of the ZAHB, MAHB, and TAHB loci on different $\MMS$ values is clearly observed. Here, it is possible to observe that the luminosity difference between the ZAHB and MAHB is larger at high $\Teff$. The luminosity difference between the ZAHB and MAHB is most pronounced at high $\Teff$, while at intermediate $\Teff$ this difference becomes almost negligible. The specific $\Teff$ at which this transition occurs depends on the chemical composition.

At the low $\Teff$ end (corresponding to high $\MHB$), the separation between these loci re-emerges, becoming more evident at higher $Y$. These findings emphasize the importance of carefully selecting the appropriate metallicity and helium abundance when comparing these evolutionary lines against observations. Interpolation of the HB tracks within a finely sampled grid is therefore crucial to avoid misrepresenting the theoretical results.

Figure~\ref{Fig:PAHBloci} also includes a comparison with ``a Bag of Stellar Tracks and Isochrones'' (BaSTI) models with $\aFe=0.4$ \citep{Hidalgo_etal2018,Pietrinferni_etal2021}, demonstrating that their ZAHB and TAHB loci (for the available $\MMS$) share similar properties to the ones from the PGPUC models. The primary differences occur at high temperatures, where the PGPUC models extend to higher $\Teff$ and lower $\MHB$. Additionally, at the lowest metallicities, the PGPUC models exhibit small differences in luminosity \textbf{($\Delta \log [L/L_\sun]\approx 0.01$)} compared to the BaSTI models for canonical He abundances, increasing to $\Delta \log (L/L_\sun)\approx 0.03$ for He-rich models. Those differences decrease when the metallicity increases. Regarding the evolutionary timescales from the ZAHB to the TAHB, Fig.~\ref{Fig:PAHBmassAge} presents the corresponding lifetimes for the same models shown in Fig.~\ref{Fig:PAHBloci}. When compared to the BaSTI models, the differences are typically $\lesssim10$~Myr across the full mass range. These discrepancies may arise from differences in the adopted input physics and/or the assumed solar heavy-element mixture.

The effects of the $Z$ value on the ZAHB locus have been well documented in the literature \citep[][among others]{Faulkner_Iben1966, Iben_Faulkner1968, Iben_Rood1970, Sweigart1987, Caputo_deglInnocenti1995, Valcarce_etal2012}; however, this is not the case for MAHB and TAHB loci. Figures~\ref{Fig:Yloci} and \ref{Fig:Zloci} show that the TAHB $\Teff$ extension is rather stable as a function of chemical composition, with the main noteworthy trend being an overall shift to higher temperatures with increasing $Z$. The ZAHB and MAHB loci, on the other hand, are much less extended in $\Teff$. This is due, on the one hand, to the fact that ZAHB and MAHB loci never reach temperatures in excess of $\log T_{\rm eff} \approx 4.55$, which, however, are easily surpassed at the TAHB level. On the other, at the very lowest metallicities, ZAHB models never reach temperatures low enough as to reach the red HB or even the instability strip region \citep[whose temperature range is about $6000\,{\rm K} \la \Teff \la 7250\,{\rm K}$;][]{CS2015}. This is consistent with the trend \citep[also observed in, for instance,][]{Sweigart1987,VandenBerg2024} of increasing minimum ZAHB temperature with decreasing metallicity. These trends imply that, at sufficiently low metallicities, the only red HB or RR Lyrae stars that can exist are those that are close to exhausting He in their cores. Extremely metal-poor ($Z \la 0.0001$) red HB stars and RR Lyrae should thus be very rare. This conclusion, originally investigated from a theoretical perspective by \citet{Cassisi_etal1996,Cassisi_etal1997}, is consistent with the recent analysis of three extremely metal-poor RR Lyrae stars by \citet{DOrazi2025}. 

Another interesting effect seen in Fig.~\ref{Fig:Yloci} that involves ZAHB and MAHB loci is that, for some chemical compositions, these loci share the same $\log (L/L_\sun)$ value at a given $\log (\Teff)$. In some cases, the MAHB locus even becomes fainter than the ZAHB locus. This is due to the fact that, at a given $\Teff$, depending on the chemical composition, the MAHB level of an HB track may be less luminous than the ZAHB level of another HB track. Indeed, it is worth noting that, for some chemical compositions, the blue loops developed by some evolutionary tracks may reach luminosities that are lower than the ZAHB. This effect can also be seen in the HB tracks presented by \citet{Sweigart1987}. 

For low $Y$, the crossing point between the ZAHB and MAHB moves to lower $\Teff$ as $Z$ increases, while when $Y$ increases, this point moves to higher $\Teff$ as $Z$ increases. For $\Teff$ values lower than the crossing point, ZAHB and MAHB loci have different behaviors depending on $Y$ and $Z$. At low $Y$, ZAHB and MAHB loci share similar $\log (\Teff)$ and $\log (L/L_\sun)$ values at high $Z$, but at very low metallicities, the MAHB locus can be fainter than the ZAHB locus. When $Y$ increases, there is a notable increase in the difference in $\log (L/L_\sun)$ between ZAHB and MAHB, causing HB stars to be spread over a larger space in the ($\log L, \log \Teff$) plane. For $\Teff$ values higher than the crossing point, there is a gradual increase in the difference in $\log (L/L_\sun)$ between ZAHB and MAHB that is less affected by $Z$ and $Y$ than happens at $\Teff$ values lower than the crossing point.

As to the impact of differences in $\MMS$, the most significant differences in ZAHB, MAHB, and TAHB loci, as is observed in Fig.~\ref{Fig:PAHBloci}, are seen primarily at higher temperatures, where a higher $\MMS$ value results in a lower luminosity at a given $\Teff$. This phenomenon is particularly pronounced at low $Z$ and high $Y$. 
The effect is produced by the larger spread in $\McHe$ values when varying $\MMS$ at the metal-poor and He-rich end (see Table~\ref{TableC}). 

Even though the ZAHB, MAHB, and TAHB loci show similar shapes in the ($\log L, \log \Teff$) plane when changing $\MMS$ for some chemical compositions, the $\MHB$ value at a given $\Teff$ does change when $\MMS$ changes. In this sense, one must be very careful when masses of HB stars are obtained from these loci without considering the most appropriate $\MMS$ value for a given age. Differences can reach as much as $6\,000$~K at low $\MHB$ values (i.e., at the hot end of the HB) for high $Z$ and low $Y$; more generally, the range in $\Teff$ values increases for lower $Z$ and higher $Y$. Although these considerations must be kept in mind when comparing observational data to theoretical models, present-day observational uncertainties in hot HB stars' $\Teff$ values \citep[e.g.,][]{Latour_etal2023} are of the same order of magnitude as the differences in $\Teff$ predicted by assuming different $\MMS$ values.

\section{Summary and conclusions}
\label{sec:conclusions}

This work introduces an extensive set of HB tracks computed with the PGPUC code. Spanning over 19,000 evolutionary tracks, this dataset accounts for a broad range of chemical compositions and progenitor masses, providing a comprehensive resource for the study of HB stars. Integrated into the PGPUC Online database, these HB tracks enable fast interpolation and an efficient comparison with observational data.

The PGPUC Online database now offers precise interpolation capabilities from the ZAHB to the TAHB, including HB tracks for different progenitor masses. This capability is critical for interpreting high-accuracy present-day observational data and serves as a very useful tool for examining the evolutionary phases of HB stars.

Our results highlight the dependency of HB stellar properties on initial conditions, such as $Z$, $Y$, and $\MMS$. In particular:
    \begin{itemize}
        \item The $\Mmin$ value decreases with increasing $Z$ and $Y$, with higher $\MMS$ leading to lower $\Mmin$.
        \item At high $Z$, the $\Mmin$ value is nearly constant for any $\MMS$ or $Y$ values. However, at low $Z$, the difference in the $\Mmin$ value when varying $\MMS$ becomes more pronounced as $Y$ increases.        
    \end{itemize}

The morphology of HB loci in the theoretical plane is strongly influenced by initial parameters. Notably, higher $\MMS$ values result in lower luminosities at a given effective temperature, emphasizing the importance of the progenitor's mass (hence age) in modeling the evolution of HB stars.

In conclusion, the expanded PGPUC Online database provides a robust framework for investigating HB stellar populations. By incorporating a wide array of initial parameters and offering publicly accessible tools for interpolation, this unique resource can be leveraged to address outstanding questions about HB morphology, the formation of blue and extreme HB stars, and how the latter are affected by changes in chemical composition and age.

\begin{acknowledgements}
We thank the referee, S. Cassisi, for a helpful report.
We also thank Yang Chen for his valuable assistance with the implementation of the PARSEC bolometric corrections into the PGPUC Online database. Support for this project is provided by ANID's FONDECYT Regular grants \#1171273 and \#1231637; ANID's Millennium Science Initiative through grants ICN12\textunderscore 009 and AIM23-0001, awarded to the Millennium Institute of Astrophysics (MAS); and ANID's Basal project FB210003. The Geryon cluster at the Centro de Astro-Ingenieria UC was extensively used for the calculations performed in this paper. BASAL CATA PFB-06, the Anillo ACT-86, FONDEQUIP AIC-57, and QUIMAL 130008 provided funding for several improvements to the Geryon cluster.
\end{acknowledgements}

\bibliographystyle{aa}
\bibliography{bibliography} 

\begin{thebibliography}{72}
\expandafter\ifx\csname natexlab\endcsname\relax\def\natexlab#1{#1}\fi

\bibitem[{{Angulo} {et~al.}(1999){Angulo}, {Arnould}, {Rayet}, {Descouvemont},
  {Baye}, {Leclercq-Willain}, {Coc}, {Barhoumi}, {Aguer}, {Rolfs}, {Kunz},
  {Hammer}, {Mayer}, {Paradellis}, {Kossionides}, {Chronidou}, {Spyrou},
  {Degl'Innocenti}, {Fiorentini}, {Ricci}, {Zavatarelli}, {Providencia},
  {Wolters}, {Soares}, {Grama}, {Rahighi}, {Shotter}, \& {Lamehi
  Rachti}}]{Angulo_etal1999}
{Angulo}, C., {Arnould}, M., {Rayet}, M., {et~al.} 1999, \nphysa, 656, 3

\bibitem[{{Bono} {et~al.}(1997{\natexlab{a}}){Bono}, {Caputo}, {Cassisi},
  {Castellani}, \& {Marconi}}]{Bono_etal1997b}
{Bono}, G., {Caputo}, F., {Cassisi}, S., {Castellani}, V., \& {Marconi}, M.
  1997{\natexlab{a}}, \apj, 489, 822

\bibitem[{{Bono} {et~al.}(1997{\natexlab{b}}){Bono}, {Caputo}, {Cassisi},
  {Castellani}, \& {Marconi}}]{Bono_etal1997a}
{Bono}, G., {Caputo}, F., {Cassisi}, S., {Castellani}, V., \& {Marconi}, M.
  1997{\natexlab{b}}, \apj, 479, 279

\bibitem[{{Caputo} \& {Raffelt}(2024)}]{Caputo_Raffelt2024}
{Caputo}, A. \& {Raffelt}, G. 2024, in 1st General Meeting and 1st Training
  School of the COST Action COSMIC WISPers, Vol.~1, 41

\bibitem[{{Caputo} \& {Degl'Innocenti}(1995)}]{Caputo_deglInnocenti1995}
{Caputo}, F. \& {Degl'Innocenti}, S. 1995, \aap, 298, 833

\bibitem[{{Carenza} {et~al.}(2025){Carenza}, {Giannotti}, {Isern}, {Mirizzi},
  \& {Straniero}}]{Carenza2025}
{Carenza}, P., {Giannotti}, M., {Isern}, J., {Mirizzi}, A., \& {Straniero}, O.
  2025, \physrep, 1117, 1

\bibitem[{{Casagrande} \& {VandenBerg}(2018)}]{Casagrande_VandenBerg2018}
{Casagrande}, L. \& {VandenBerg}, D.~A. 2018, \mnras, 475, 5023

\bibitem[{{Cassisi} {et~al.}(1997){Cassisi}, {Castellani}, \&
  {Castellani}}]{Cassisi_etal1997}
{Cassisi}, S., {Castellani}, M., \& {Castellani}, V. 1997, \aap, 317, 108

\bibitem[{{Cassisi} {et~al.}(1996){Cassisi}, {Castellani}, \&
  {Tornambe}}]{Cassisi_etal1996}
{Cassisi}, S., {Castellani}, V., \& {Tornambe}, A. 1996, \apj, 459, 298

\bibitem[{{Cassisi} {et~al.}(2007){Cassisi}, {Potekhin}, {Pietrinferni},
  {Catelan}, \& {Salaris}}]{Cassisi_etal2007}
{Cassisi}, S., {Potekhin}, A.~Y., {Pietrinferni}, A., {Catelan}, M., \&
  {Salaris}, M. 2007, \apj, 661, 1094

\bibitem[{{Cassisi} {et~al.}(2003){Cassisi}, {Salaris}, \&
  {Irwin}}]{Cassisi_etal2003}
{Cassisi}, S., {Salaris}, M., \& {Irwin}, A.~W. 2003, \apj, 588, 862

\bibitem[{{Castellani} {et~al.}(1994){Castellani}, {Castellani}, {Pulone}, \&
  {Tornambe}}]{Castellani_etal1994}
{Castellani}, M., {Castellani}, V., {Pulone}, L., \& {Tornambe}, A. 1994, \aap,
  282, 771

\bibitem[{{Catelan}(2009)}]{Catelan2009}
{Catelan}, M. 2009, \apss, 320, 261

\bibitem[{{Catelan} {et~al.}(2009){Catelan}, {Grundahl}, {Sweigart},
  {Valcarce}, \& {Cort{\'e}s}}]{Catelan_etal2009}
{Catelan}, M., {Grundahl}, F., {Sweigart}, A.~V., {Valcarce}, A.~A.~R., \&
  {Cort{\'e}s}, C. 2009, \apjl, 695, L97

\bibitem[{{Catelan} \& {Smith}(2015)}]{CS2015}
{Catelan}, M. \& {Smith}, H.~A. 2015, {Pulsating Stars (Wiley-VCH, Weinheim)}

\bibitem[{{Chen} {et~al.}(2019){Chen}, {Girardi}, {Fu}, {Bressan}, {Aringer},
  {Dal Tio}, {Pastorelli}, {Marigo}, {Costa}, \& {Zhang}}]{Chen_etal2019}
{Chen}, Y., {Girardi}, L., {Fu}, X., {et~al.} 2019, \aap, 632, A105

\bibitem[{{Demarque} \& {Mengel}(1972)}]{Demarque_Mengel1972}
{Demarque}, P. \& {Mengel}, J.~G. 1972, \apj, 171, 583

\bibitem[{{DeWitt} {et~al.}(1973){DeWitt}, {Graboske}, \&
  {Cooper}}]{Dewitt1973}
{DeWitt}, H.~E., {Graboske}, H.~C., \& {Cooper}, M.~S. 1973, \apj, 181, 439

\bibitem[{{D'Orazi} {et~al.}(2025){D'Orazi}, {Braga}, {Bono}, {Fabrizio},
  {Fiorentino}, {Storm}, {Pietrinferni}, {Sneden}, {S{\'a}nchez-Benavente},
  {Monelli}, {Sestito}, {J{\"o}nsson}, {Buder}, {Bobrick}, {Iorio},
  {Matsunaga}, {Marconi}, {Marengo}, {Mart{\'\i}nez-V{\'a}zquez}, {Mullen},
  {Takayama}, {Testa}, {Cusano}, \& {Crestani}}]{DOrazi2025}
{D'Orazi}, V., {Braga}, V., {Bono}, G., {et~al.} 2025, \aap, 694, A158

\bibitem[{{Dorman}(1992)}]{Dorman1992}
{Dorman}, B. 1992, \apjs, 80, 701

\bibitem[{{Dotter} {et~al.}(2007){Dotter}, {Chaboyer}, {Jevremovi{\'c}},
  {Baron}, {Ferguson}, {Sarajedini}, \& {Anderson}}]{Dotter_etal2007}
{Dotter}, A., {Chaboyer}, B., {Jevremovi{\'c}}, D., {et~al.} 2007, \aj, 134,
  376

\bibitem[{{Dotter} {et~al.}(2008){Dotter}, {Chaboyer}, {Jevremovi{\'c}},
  {Kostov}, {Baron}, \& {Ferguson}}]{Dotter_etal2008}
{Dotter}, A., {Chaboyer}, B., {Jevremovi{\'c}}, D., {et~al.} 2008, \apjs, 178,
  89

\bibitem[{{Eggleton}(1968)}]{Eggleton1968}
{Eggleton}, P.~P. 1968, \mnras, 140, 387

\bibitem[{{Faulkner}(1966)}]{Faulkner1966}
{Faulkner}, J. 1966, \apj, 144, 978

\bibitem[{{Faulkner} \& {Iben}(1966)}]{Faulkner_Iben1966}
{Faulkner}, J. \& {Iben}, Jr., I. 1966, \apj, 144, 995

\bibitem[{{Ferguson} {et~al.}(2005){Ferguson}, {Alexander}, {Allard}, {Barman},
  {Bodnarik}, {Hauschildt}, {Heffner-Wong}, \& {Tamanai}}]{Ferguson_etal2005}
{Ferguson}, J.~W., {Alexander}, D.~R., {Allard}, F., {et~al.} 2005, \apj, 623,
  585

\bibitem[{{Graboske} {et~al.}(1973){Graboske}, {Dewitt}, {Grossman}, \&
  {Cooper}}]{Graboske_etal1973}
{Graboske}, H.~C., {Dewitt}, H.~E., {Grossman}, A.~S., \& {Cooper}, M.~S. 1973,
  \apj, 181, 457

\bibitem[{{Grevesse} \& {Sauval}(1998)}]{Grevesse_Sauval1998}
{Grevesse}, N. \& {Sauval}, A.~J. 1998, \ssr, 85, 161

\bibitem[{{Gross}(1973)}]{Gross1973}
{Gross}, P.~G. 1973, \mnras, 164, 65

\bibitem[{{Haft} {et~al.}(1994){Haft}, {Raffelt}, \& {Weiss}}]{Haft_etal1994}
{Haft}, M., {Raffelt}, G., \& {Weiss}, A. 1994, \apj, 425, 222

\bibitem[{{H{\"a}rm} \& {Schwarzschild}(1966)}]{Harm_Schwarzschild1966}
{H{\"a}rm}, R. \& {Schwarzschild}, M. 1966, \apj, 145, 496

\bibitem[{{Hidalgo} {et~al.}(2018){Hidalgo}, {Pietrinferni}, {Cassisi},
  {Salaris}, {Mucciarelli}, {Savino}, {Aparicio}, {Silva Aguirre}, \&
  {Verma}}]{Hidalgo_etal2018}
{Hidalgo}, S.~L., {Pietrinferni}, A., {Cassisi}, S., {et~al.} 2018, \apj, 856,
  125

\bibitem[{{Iben}(1964)}]{Iben1964}
{Iben}, Jr., I. 1964, \apj, 140, 1631

\bibitem[{{Iben} \& {Faulkner}(1968)}]{Iben_Faulkner1968}
{Iben}, Jr., I. \& {Faulkner}, J. 1968, \apj, 153, 101

\bibitem[{{Iben} \& {Rood}(1970)}]{Iben_Rood1970}
{Iben}, Jr., I. \& {Rood}, R.~T. 1970, \apj, 161, 587

\bibitem[{{Iglesias} \& {Rogers}(1996)}]{Iglesias_Rogers1996}
{Iglesias}, C.~A. \& {Rogers}, F.~J. 1996, \apj, 464, 943

\bibitem[{{Irwin}(2007)}]{Irwin2007}
{Irwin}, A.~W. 2007, http://freeeos.sourceforge.net

\bibitem[{{Krishna Swamy}(1966)}]{Krishna_Swamy1966}
{Krishna Swamy}, K.~S. 1966, \apj, 145, 174

\bibitem[{{Kunz} {et~al.}(2002){Kunz}, {Fey}, {Jaeger}, {Mayer}, {Hammer},
  {Staudt}, {Harissopulos}, \& {Paradellis}}]{Kunz_etal2002}
{Kunz}, R., {Fey}, M., {Jaeger}, M., {et~al.} 2002, \apj, 567, 643

\bibitem[{{Latour} {et~al.}(2023){Latour}, {H{\"a}mmerich}, {Dorsch}, {Heber},
  {Husser}, {Kamman}, {Dreizler}, \& {Brinchmann}}]{Latour_etal2023}
{Latour}, M., {H{\"a}mmerich}, S., {Dorsch}, M., {et~al.} 2023, \aap, 677, A86

\bibitem[{{Pietrinferni} {et~al.}(2004){Pietrinferni}, {Cassisi}, {Salaris}, \&
  {Castelli}}]{Pietrinferni_etal2004}
{Pietrinferni}, A., {Cassisi}, S., {Salaris}, M., \& {Castelli}, F. 2004, \apj,
  612, 168

\bibitem[{{Pietrinferni} {et~al.}(2013){Pietrinferni}, {Cassisi}, {Salaris}, \&
  {Hidalgo}}]{Pietrinferni_etal2013}
{Pietrinferni}, A., {Cassisi}, S., {Salaris}, M., \& {Hidalgo}, S. 2013, \aap,
  558, A46

\bibitem[{{Pietrinferni} {et~al.}(2021){Pietrinferni}, {Hidalgo}, {Cassisi},
  {Salaris}, {Savino}, {Mucciarelli}, {Verma}, {Silva Aguirre}, {Aparicio}, \&
  {Ferguson}}]{Pietrinferni_etal2021}
{Pietrinferni}, A., {Hidalgo}, S., {Cassisi}, S., {et~al.} 2021, \apj, 908, 102

\bibitem[{{Renzini}(1977)}]{Renzini1977}
{Renzini}, A. 1977, in Saas-Fee Advanced Course 7: Advanced Stages in Stellar
  Evolution, ed. P.~{Bouvier} \& A.~{Maeder}, 151

\bibitem[{{Rood}(1973)}]{Rood1973}
{Rood}, R.~T. 1973, \apj, 184, 815

\bibitem[{{Salaris} \& {Cassisi}(2005)}]{Salaris_Cassisi2005}
{Salaris}, M. \& {Cassisi}, S. 2005, {Evolution of Stars and Stellar
  Populations}

\bibitem[{{Salpeter}(1954)}]{Salpeter1954}
{Salpeter}, E.~E. 1954, Australian Journal of Physics, 7, 373

\bibitem[{{Schr{\"o}der} \& {Cuntz}(2005)}]{Schroder_Cuntz2005}
{Schr{\"o}der}, K.~P. \& {Cuntz}, M. 2005, \apjl, 630, L73

\bibitem[{{Schwarzschild} \& {H{\"a}rm}(1965)}]{Schwarzschild_Harm1965}
{Schwarzschild}, M. \& {H{\"a}rm}, R. 1965, \apj, 142, 855

\bibitem[{{Serenelli} {et~al.}(2017){Serenelli}, {Weiss}, {Cassisi}, {Salaris},
  \& {Pietrinferni}}]{Serenelli_etal2017}
{Serenelli}, A., {Weiss}, A., {Cassisi}, S., {Salaris}, M., \& {Pietrinferni},
  A. 2017, \aap, 606, A33

\bibitem[{{Straniero} {et~al.}(2020){Straniero}, {Pallanca}, {Dalessandro},
  {Dom{\'\i}nguez}, {Ferraro}, {Giannotti}, {Mirizzi}, \&
  {Piersanti}}]{Straniero_etal2020}
{Straniero}, O., {Pallanca}, C., {Dalessandro}, E., {et~al.} 2020, \aap, 644,
  A166

\bibitem[{{Sweigart}(1971)}]{Sweigart1971}
{Sweigart}, A.~V. 1971, \apj, 168, 79

\bibitem[{{Sweigart}(1973)}]{Sweigart1973}
{Sweigart}, A.~V. 1973, \aap, 24, 459

\bibitem[{{Sweigart}(1987)}]{Sweigart1987}
{Sweigart}, A.~V. 1987, \apjs, 65, 95

\bibitem[{{Sweigart}(1994)}]{Sweigart1994}
{Sweigart}, A.~V. 1994, \apj, 426, 612

\bibitem[{{Sweigart}(1997)}]{Sweigart1997}
{Sweigart}, A.~V. 1997, \apjl, 474, L23

\bibitem[{{Sweigart} \& {Catelan}(1998)}]{Sweigart_Catelan1998}
{Sweigart}, A.~V. \& {Catelan}, M. 1998, \apjl, 501, L63

\bibitem[{{Sweigart} \& {Demarque}(1972)}]{Sweigart_Demarque1972}
{Sweigart}, A.~V. \& {Demarque}, P. 1972, \aap, 20, 445

\bibitem[{{Sweigart} \& {Gross}(1974)}]{Sweigart_Gross1974}
{Sweigart}, A.~V. \& {Gross}, P.~G. 1974, \apj, 190, 101

\bibitem[{{Sweigart} \& {Gross}(1976)}]{Sweigart_Gross1976}
{Sweigart}, A.~V. \& {Gross}, P.~G. 1976, \apjs, 32, 367

\bibitem[{{Sweigart} \& {Gross}(1978)}]{Sweigart_Gross1978}
{Sweigart}, A.~V. \& {Gross}, P.~G. 1978, \apjs, 36, 405

\bibitem[{{Valcarce}(2011)}]{Valcarce2011}
{Valcarce}, A. A.~R. 2011, {Study of the Helium Enrichment in Globular Clusters
  (Ph.D. Thesis, Pontificia Universidad Católica de Chile)}

\bibitem[{{Valcarce} {et~al.}(2016){Valcarce}, {Catelan}, {Alonso-Garc{\'\i}a},
  {Contreras Ramos}, \& {Alves}}]{Valcarce_etal2016}
{Valcarce}, A.~A.~R., {Catelan}, M., {Alonso-Garc{\'\i}a}, J., {Contreras
  Ramos}, R., \& {Alves}, S. 2016, \aap, 589, A126

\bibitem[{{Valcarce} {et~al.}(2014){Valcarce}, {Catelan}, {Alonso-Garc{\'\i}a},
  {Cort{\'e}s}, \& {De Medeiros}}]{Valcarce_etal2014}
{Valcarce}, A.~A.~R., {Catelan}, M., {Alonso-Garc{\'\i}a}, J., {Cort{\'e}s},
  C., \& {De Medeiros}, J.~R. 2014, \apj, 782, 85

\bibitem[{{Valcarce} {et~al.}(2012){Valcarce}, {Catelan}, \&
  {Sweigart}}]{Valcarce_etal2012}
{Valcarce}, A.~A.~R., {Catelan}, M., \& {Sweigart}, A.~V. 2012, \aap, 547, A5

\bibitem[{{VandenBerg}(1992)}]{Vandenberg1992}
{VandenBerg}, D.~A. 1992, \apj, 391, 685

\bibitem[{{VandenBerg}(2024)}]{VandenBerg2024}
{VandenBerg}, D.~A. 2024, \mnras, 527, 6888

\bibitem[{{VandenBerg} {et~al.}(2006){VandenBerg}, {Bergbusch}, \&
  {Dowler}}]{VandenBerg_etal2006}
{VandenBerg}, D.~A., {Bergbusch}, P.~A., \& {Dowler}, P.~D. 2006, \apjs, 162,
  375

\bibitem[{{VandenBerg} {et~al.}(2000){VandenBerg}, {Swenson}, {Rogers},
  {Iglesias}, \& {Alexander}}]{VandenBerg2000}
{VandenBerg}, D.~A., {Swenson}, F.~J., {Rogers}, F.~J., {Iglesias}, C.~A., \&
  {Alexander}, D.~R. 2000, \apj, 532, 430

\bibitem[{{Viaux} {et~al.}(2013{\natexlab{a}}){Viaux}, {Catelan}, {Stetson},
  {Raffelt}, {Redondo}, {Valcarce}, \& {Weiss}}]{Viaux_etal2013b}
{Viaux}, N., {Catelan}, M., {Stetson}, P.~B., {et~al.} 2013{\natexlab{a}},
  \prl, 111, 231301

\bibitem[{{Viaux} {et~al.}(2013{\natexlab{b}}){Viaux}, {Catelan}, {Stetson},
  {Raffelt}, {Redondo}, {Valcarce}, \& {Weiss}}]{Viaux_etal2013a}
{Viaux}, N., {Catelan}, M., {Stetson}, P.~B., {et~al.} 2013{\natexlab{b}},
  \aap, 558, A12

\bibitem[{{Zoccali} {et~al.}(2000){Zoccali}, {Cassisi}, {Bono}, {Piotto},
  {Rich}, \& {Djorgovski}}]{Zoccali_etal2000}
{Zoccali}, M., {Cassisi}, S., {Bono}, G., {et~al.} 2000, \apj, 538, 289

\end{thebibliography}

\begin{appendix}
\onecolumn
\section{Electron screening and the RGB tip luminosity}
\label{sec:appendix}

\citet{Valcarce_etal2012} carried out a comparison between the results obtained with PGPUC and five other SECs. PGPUC's predecessor, the Princeton-Goddard (PG) SEC (see their Sect.~2 for extensive references), was also added to that analysis. One noteworthy result of these comparisons, well illustrated in their Fig.~3, is the realization that, for similar input parameters, there is a fairly wide range in predicted RGB tip luminosities. In particular, PG and PGPUC predict the faintest RGB tips. A comparison between PG and Victoria results had previously been carried out by \citet{VandenBerg2000}, who found that PG predicted similar but slightly fainter RGB tips and ZAHB loci, in spite of very similar $\McHe$ values.  

The reason why this happens is unclear, and explaining it is well beyond the scope of the present paper. However, it has been suggested in the literature \citep{Serenelli_etal2017, Straniero_etal2020, Caputo_Raffelt2024, Carenza2025} that this can be traced in part to the fact that PGPUC presumably uses Salpeter’s formulation of weak screening of nuclear reactions \citep{Salpeter1954}, leading to inaccurate predictions of the position of the RGB tip luminosity, as compared to other codes that use more up-to-date recipes for electron screening.  

This explanation is not correct. As explained in \citet[][Sect.~II.b]{Sweigart_Gross1978}, the \citet{Dewitt1973} and \citet{Graboske_etal1973} prescriptions were adopted early on in the development of the PG code to account for electron screening. PGPUC is an updated version of the PG SEC \citep{Valcarce_etal2012}. While the input physics in PGPUC has been extensively updated since \citet{Sweigart_Gross1978}, it still relies, as do other state-of-the-art SECs, on \citet{Dewitt1973} and \citet{Graboske_etal1973} for the treatment of electron screening. 

The confusion appears to stem from a misinterpretation of \citet{Viaux_etal2013b,Viaux_etal2013a}, where PGPUC models were used to carry out a detailed study of the RGB tip position in the GC M5 (NGC\,5904). As the only reference that \citeauthor{Viaux_etal2013a} cite to introduce the subject of screening is \citet{Salpeter1954}, this may have led to the incorrect impression that \citeauthor{Salpeter1954}'s prescriptions were used in the actual calculations, when in fact they were not. The fact that electron screening is not specifically addressed in either \citet{Valcarce2011} or \citet{Valcarce_etal2012} may have contributed to the confusion. 

In conclusion, whatever the ultimate explanation for the differences in RGB tip position between PG/PGPUC and other codes, they are most certainly not due to adoption of the \citet{Salpeter1954} prescription for electron screening of nuclear reactions in either PG or PGPUC. A deeper analysis will be required to fully understand the reason why different SECs still predict significantly different RGB tip luminosities. 

\section{Supplementary table}
\label{sec:extratables}

Table~\ref{TableC} lists key physical properties of stellar models at the RGB tip and near the RR Lyrae instability strip. For each combination of chemical composition $(Y, \, Z)$ and progenitor mass $\MMS$, we report the helium core mass at the RGB tip ($\McHe$), the amount of additional helium in the envelope due to the first dredge-up ($\DYRGB$), the stellar age at the RGB tip (from the zero-age MS), and the luminosity ($\lLRR$) and stellar mass ($\MRR$) of the ZAHB model with $\log\Teff = 3.83$ (which is a common reference temperature near the center of the RR Lyrae instability strip). For each $Z$ value, solar-scaled and $\alpha$-enhanced compositions are considered ($[\alpha/{\rm Fe}] = 0.0$ and 0.3, respectively). At very low metallicities, ZAHB models may not reach this temperature, in which case the corresponding $\lLRR$ and $\MRR$ values are not provided. $\FeH$ values are calculated assuming $Y=0.245$.

\onecolumn
{\fontsize{7.5pt}{9pt}\selectfont\setlength{\tabcolsep}{3pt}
\renewcommand\arraystretch{1.1}
\begin{longtable}{ccccccc c ccccccc}
\caption{Stellar model parameters for different \label{TableC}
$Z$, $Y$, and $\MMS$ values.}\\  
\toprule
\toprule
$Y$ & $\MMS$ & $\McHe$ & $\DYRGB$ & Age   & $\lLRR$ & $\MRR$ & &
$Y$ & $\MMS$ & $\McHe$ & $\DYRGB$ & Age   & $\lLRR$ & $\MRR$ \\
    & ($\Mo$)& ($\Mo$) &          & (Myr) &         & ($\Mo$) & &
    & ($\Mo$)& ($\Mo$) &          & (Myr) &         & ($\Mo$) \\
\midrule
\endfirsthead

\caption{Continued.} \\
\toprule
$Y$ & $\MMS$ & $\McHe$ & $\DYRGB$ & Age   & $\lLRR$ & $\MRR$ & &
$Y$ & $\MMS$ & $\McHe$ & $\DYRGB$ & Age   & $\lLRR$ & $\MRR$ \\
    & ($\Mo$)& (

$\Mo$) &          & (Myr) &         & ($\Mo$) & &
    & ($\Mo$)& ($\Mo$) &          & (Myr) &         & ($\Mo$) \\
\midrule
\endhead

\midrule \multicolumn{15}{r}{\textit{Continued on next page}} \\
\endfoot

\bottomrule
\endlastfoot

\multicolumn{15}{c}{$Z = 0.00001$} \\
\midrule[\cmidrulewidth]
\multicolumn{7}{c}{$\FeH = -3.24, \aFe = 0.00$} & & \multicolumn{7}{c}{$\FeH = -3.47, \aFe = 0.30$} \\
\cmidrule{1-7} \cmidrule{9-15}
0.230 & 0.700 & 0.5165 & 0.0002 & 21866  & --    & --    & & 0.230 & 0.700 & 0.5167 & 0.0002 & 21867  & --    & --    \\
      & 0.800 & 0.5138 & 0.0025 & 13587  & --    & --    & &       & 0.800 & 0.5140 & 0.0025 & 13589  & --    & --    \\
      & 0.900 & 0.5099 & 0.0066 & 9005   & --    & --    & &       & 0.900 & 0.5101 & 0.0065 & 9007   & --    & --    \\
0.245 & 0.700 & 0.5126 & 0.0003 & 19989  & --    & --    & & 0.245 & 0.700 & 0.5128 & 0.0003 & 19990  & --    & --    \\
      & 0.800 & 0.5096 & 0.0028 & 12428  & --    & --    & &       & 0.800 & 0.5098 & 0.0028 & 12430  & --    & --    \\
      & 0.900 & 0.5052 & 0.0068 & 8246   & --    & --    & &       & 0.900 & 0.5054 & 0.0067 & 8248   & --    & --    \\
0.270 & 0.700 & 0.5060 & 0.0005 & 17162  & --    & --    & & 0.270 & 0.700 & 0.5062 & 0.0005 & 17164  & --    & --    \\
      & 0.800 & 0.5023 & 0.0032 & 10686  & --    & --    & &       & 0.800 & 0.5025 & 0.0031 & 10688  & --    & --    \\
      & 0.900 & 0.4969 & 0.0070 & 7107   & --    & --    & &       & 0.900 & 0.4971 & 0.0069 & 7109   & --    & --    \\
0.295 & 0.700 & 0.4992 & 0.0007 & 14685  & --    & --    & & 0.295 & 0.700 & 0.4994 & 0.0007 & 14686  & --    & --    \\
      & 0.800 & 0.4946 & 0.0035 & 9162   & --    & --    & &       & 0.800 & 0.4948 & 0.0035 & 9163   & --    & --    \\
      & 0.900 & 0.4879 & 0.0070 & 6109   & --    & --    & &       & 0.900 & 0.4880 & 0.0070 & 6110   & --    & --    \\
0.320 & 0.700 & 0.4921 & 0.0009 & 12523  & --    & --    & & 0.320 & 0.700 & 0.4922 & 0.0009 & 12524  & --    & --    \\
      & 0.800 & 0.4865 & 0.0038 & 7833   & --    & --    & &       & 0.800 & 0.4867 & 0.0037 & 7833   & --    & --    \\
      & 0.900 & 0.4775 & 0.0069 & 5236   & --    & --    & &       & 0.900 & 0.4776 & 0.0068 & 5236   & --    & --    \\
0.345 & 0.700 & 0.4846 & 0.0012 & 10619  & --    & --    & & 0.345 & 0.700 & 0.4848 & 0.0011 & 10644  & --    & --    \\
      & 0.800 & 0.4775 & 0.0039 & 6658   & --    & --    & &       & 0.800 & 0.4778 & 0.0039 & 6677   & --    & --    \\
      & 0.900 & 0.4655 & 0.0068 & 4474   & --    & --    & &       & 0.900 & 0.4657 & 0.0067 & 4474   & --    & --    \\
0.370 & 0.700 & 0.4767 & 0.0014 & 9017   & --    & --    & & 0.370 & 0.700 & 0.4769 & 0.0013 & 9017   & --    & --    \\
      & 0.800 & 0.4678 & 0.0040 & 5673   & --    & --    & &       & 0.800 & 0.4679 & 0.0039 & 5673   & --    & --    \\
      & 0.900 & 0.4516 & 0.0065 & 3800   & --    & --    & &       & 0.900 & 0.4519 & 0.0065 & 3805   & --    & --    \\
\midrule

\multicolumn{15}{c}{$Z = 0.00016$} \\
\midrule[\cmidrulewidth]
\multicolumn{7}{c}{$\FeH = -2.04, \aFe = 0.00$} & & \multicolumn{7}{c}{$\FeH = -2.26, \aFe = 0.30$} \\
\cmidrule{1-7} \cmidrule{9-15}
0.230 & 0.700 & 0.5004 & 0.0017 & 22090  & --    & --    & & 0.230 & 0.700 & 0.5007 & 0.0017 & 22102   & --    & --    \\
      & 0.800 & 0.4970 & 0.0066 & 13698  & 1.699 & 0.783 & &       & 0.800 & 0.4973 & 0.0064 & 13705   & 1.703 & 0.790 \\
      & 0.900 & 0.4933 & 0.0116 &  9052  & 1.697 & 0.776 & &       & 0.900 & 0.4935 & 0.0114 &  9055   & 1.701 & 0.784 \\
0.245 & 0.700 & 0.4967 & 0.0020 & 20185  & --    & --    & & 0.245 & 0.700 & 0.4970 & 0.0019 & 20195   & --    & --    \\
      & 0.800 & 0.4932 & 0.0069 & 12521  & 1.718 & 0.783 & &       & 0.800 & 0.4934 & 0.0067 & 12526   & 1.722 & 0.790 \\
      & 0.900 & 0.4891 & 0.0116 &  8286  & 1.714 & 0.775 & &       & 0.900 & 0.4893 & 0.0114 &  8286   & 1.718 & 0.783 \\
0.270 & 0.700 & 0.4905 & 0.0025 & 17318  & --    & --    & & 0.270 & 0.700 & 0.4907 & 0.0024 & 17326   & --    & --    \\
      & 0.800 & 0.4865 & 0.0073 & 10754  & 1.749 & 0.780 & &       & 0.800 & 0.4867 & 0.0071 & 10757   & 1.753 & 0.788 \\
      & 0.900 & 0.4820 & 0.0114 &  7130  & 1.741 & 0.770 & &       & 0.900 & 0.4821 & 0.0112 &  7132   & 1.746 & 0.778 \\
0.295 & 0.700 & 0.4841 & 0.0030 & 14805  & --    & --    & & 0.295 & 0.700 & 0.4844 & 0.0029 & 14810   & --    & --    \\
      & 0.800 & 0.4797 & 0.0074 &  9209  & 1.778 & 0.774 & &       & 0.800 & 0.4799 & 0.0073 &  9211   & 1.783 & 0.782 \\
      & 0.900 & 0.4745 & 0.0111 &  6122  & 1.768 & 0.761 & &       & 0.900 & 0.4746 & 0.0110 &  6121   & 1.772 & 0.769 \\
0.320 & 0.700 & 0.4775 & 0.0034 & 12614  & --    & --    & & 0.320 & 0.700 & 0.4777 & 0.0032 & 12616   & --    & --    \\
      & 0.800 & 0.4725 & 0.0074 &  7864  & 1.807 & 0.765 & &       & 0.800 & 0.4727 & 0.0073 &  7864   & 1.811 & 0.772 \\
      & 0.900 & 0.4665 & 0.0107 &  5241  & 1.792 & 0.749 & &       & 0.900 & 0.4665 & 0.0106 &  5240   & 1.796 & 0.756 \\
0.345 & 0.700 & 0.4708 & 0.0037 & 10710  & --    & --    & & 0.345 & 0.700 & 0.4710 & 0.0036 & 10710   & --    & --    \\
      & 0.800 & 0.4651 & 0.0073 &  6695  & 1.834 & 0.752 & &       & 0.800 & 0.4652 & 0.0072 &  6694   & 1.838 & 0.760 \\
      & 0.900 & 0.4579 & 0.0102 &  4473  & 1.813 & 0.732 & &       & 0.900 & 0.4578 & 0.0100 &  4472   & 1.817 & 0.739 \\
0.370 & 0.700 & 0.4639 & 0.0039 &  9062  & --    & --    & & 0.370 & 0.700 & 0.4640 & 0.0038 &  9061   & --    & --    \\
      & 0.800 & 0.4572 & 0.0070 &  5682  & 1.859 & 0.737 & &       & 0.800 & 0.4573 & 0.0070 &  5681   & 1.864 & 0.744 \\
      & 0.900 & 0.4484 & 0.0095 &  3806  & 1.832 & 0.714 & &       & 0.900 & 0.4483 & 0.0095 &  3804   & 1.835 & 0.720 \\
\midrule

\multicolumn{15}{c}{$Z = 0.00028$} \\
\midrule[\cmidrulewidth]
\multicolumn{7}{c}{$\FeH = -1.79, \aFe = 0.00$} & & \multicolumn{7}{c}{$\FeH = -2.02, \aFe = 0.30$} \\
\cmidrule{1-7} \cmidrule{9-15}
0.230 & 0.700 & 0.4973 & 0.0024 & 22384   & --    & --    & & 0.230 & 0.700 & 0.4975 & 0.0023 & 22395  & --    & --    \\
      & 0.800 & 0.4939 & 0.0078 & 13873   & 1.669 & 0.730 & &       & 0.800 & 0.4941 & 0.0075 & 13878  & 1.673 & 0.736 \\
      & 0.900 & 0.4903 & 0.0128 &  9156   & 1.667 & 0.725 & &       & 0.900 & 0.4905 & 0.0125 &  9158  & 1.670 & 0.731 \\
0.245 & 0.700 & 0.4937 & 0.0027 & 20449   & --    & --    & & 0.245 & 0.700 & 0.4939 & 0.0026 & 20457  & --    & --    \\
      & 0.800 & 0.4901 & 0.0080 & 12676   & 1.687 & 0.728 & &       & 0.800 & 0.4903 & 0.0078 & 12680  & 1.691 & 0.735 \\
      & 0.900 & 0.4863 & 0.0127 &  8376   & 1.684 & 0.722 & &       & 0.900 & 0.4865 & 0.0125 &  8377  & 1.688 & 0.729 \\
0.270 & 0.700 & 0.4875 & 0.0032 & 17538   & --    & --    & & 0.270 & 0.700 & 0.4878 & 0.0031 & 17543  & --    & --    \\
      & 0.800 & 0.4837 & 0.0083 & 10881   & 1.717 & 0.723 & &       & 0.800 & 0.4839 & 0.0081 & 10883  & 1.721 & 0.730 \\
      & 0.900 & 0.4795 & 0.0125 &  7205   & 1.711 & 0.716 & &       & 0.900 & 0.4796 & 0.0123 &  7204  & 1.715 & 0.722 \\
0.295 & 0.700 & 0.4813 & 0.0038 & 14988   & --    & --    & & 0.295 & 0.700 & 0.4815 & 0.0036 & 14991  & --    & --    \\
      & 0.800 & 0.4772 & 0.0085 &  9313   & 1.747 & 0.717 & &       & 0.800 & 0.4773 & 0.0083 &  9313  & 1.751 & 0.723 \\
      & 0.900 & 0.4724 & 0.0121 &  6181   & 1.738 & 0.707 & &       & 0.900 & 0.4724 & 0.0119 &  6180  & 1.742 & 0.713 \\
0.320 & 0.700 & 0.4750 & 0.0042 & 12764   & --    & --    & & 0.320 & 0.700 & 0.4752 & 0.0041 & 12763  & --    & --    \\
      & 0.800 & 0.4703 & 0.0084 &  7946   & 1.776 & 0.707 & &       & 0.800 & 0.4704 & 0.0082 &  7945  & 1.780 & 0.714 \\
      & 0.900 & 0.4649 & 0.0116 &  5288   & 1.765 & 0.696 & &       & 0.900 & 0.4649 & 0.0114 &  5286  & 1.768 & 0.702 \\
0.345 & 0.700 & 0.4685 & 0.0045 & 10833   & --    & --    & & 0.345 & 0.700 & 0.4686 & 0.0044 & 10826  & --    & --    \\
      & 0.800 & 0.4632 & 0.0082 &  6761   & 1.805 & 0.696 & &       & 0.800 & 0.4633 & 0.0081 &  6758  & 1.809 & 0.702 \\
      & 0.900 & 0.4570 & 0.0110 &  4511   & 1.790 & 0.684 & &       & 0.900 & 0.4569 & 0.0108 &  4508  & 1.793 & 0.688 \\
0.370 & 0.700 & 0.4618 & 0.0046 &  9160   & 1.849 & 0.698 & & 0.370 & 0.700 & 0.4619 & 0.0045 &  9155  & --    & --    \\
      & 0.800 & 0.4558 & 0.0079 &  5734   & 1.833 & 0.684 & &       & 0.800 & 0.4558 & 0.0077 &  5730  & 1.840 & 0.688 \\
      & 0.900 & 0.4484 & 0.0103 &  3837   & 1.813 & 0.669 & &       & 0.900 & 0.4483 & 0.0102 &  3833  & 1.815 & 0.673 \\
\midrule

\multicolumn{15}{c}{$Z = 0.00051$} \\
\midrule[\cmidrulewidth]
\multicolumn{7}{c}{$\FeH = -1.53, \aFe = 0.00$} & & \multicolumn{7}{c}{$\FeH = -1.76, \aFe = 0.30$} \\
\cmidrule{1-7} \cmidrule{9-15}
0.230 & 0.700 & 0.4941 & 0.0031 & 22978   & 1.639 & 0.690 & & 0.230 & 0.700 & 0.4943 & 0.0030 & 22981  & 1.643 & 0.696 \\
      & 0.800 & 0.4908 & 0.0091 & 14229   & 1.640 & 0.686 & &       & 0.800 & 0.4910 & 0.0088 & 14225  & 1.644 & 0.692 \\
      & 0.900 & 0.4875 & 0.0142 &  9375   & 1.639 & 0.682 & &       & 0.900 & 0.4876 & 0.0140 &  9367  & 1.643 & 0.688 \\
0.245 & 0.700 & 0.4906 & 0.0036 & 20983   & 1.658 & 0.688 & & 0.245 & 0.700 & 0.4908 & 0.0034 & 20983  & 1.662 & 0.694 \\
      & 0.800 & 0.4871 & 0.0093 & 12997   & 1.658 & 0.684 & &       & 0.800 & 0.4873 & 0.0091 & 12991  & 1.662 & 0.689 \\
      & 0.900 & 0.4837 & 0.0140 &  8571   & 1.656 & 0.679 & &       & 0.900 & 0.4838 & 0.0138 &  8562  & 1.660 & 0.684 \\
0.270 & 0.700 & 0.4846 & 0.0041 & 17985   & 1.690 & 0.683 & & 0.270 & 0.700 & 0.4849 & 0.0040 & 17981  & 1.695 & 0.689 \\
      & 0.800 & 0.4810 & 0.0096 & 11145   & 1.689 & 0.678 & &       & 0.800 & 0.4811 & 0.0094 & 11137  & 1.693 & 0.683 \\
      & 0.900 & 0.4772 & 0.0137 &  7365   & 1.684 & 0.672 & &       & 0.900 & 0.4773 & 0.0135 &  7355  & 1.689 & 0.678 \\
0.295 & 0.700 & 0.4786 & 0.0047 & 15360   & 1.723 & 0.677 & & 0.295 & 0.700 & 0.4787 & 0.0045 & 15353  & 1.727 & 0.682 \\
      & 0.800 & 0.4747 & 0.0096 &  9529   & 1.720 & 0.671 & &       & 0.800 & 0.4748 & 0.0094 &  9519  & 1.724 & 0.676 \\
      & 0.900 & 0.4706 & 0.0132 &  6312   & 1.713 & 0.664 & &       & 0.900 & 0.4705 & 0.0130 &  6302  & 1.717 & 0.669 \\
0.320 & 0.700 & 0.4725 & 0.0052 & 13070   & 1.756 & 0.669 & & 0.320 & 0.700 & 0.4726 & 0.0050 & 13061  & 1.760 & 0.674 \\
      & 0.800 & 0.4683 & 0.0095 &  8122   & 1.751 & 0.662 & &       & 0.800 & 0.4683 & 0.0093 &  8112  & 1.755 & 0.667 \\
      & 0.900 & 0.4637 & 0.0126 &  5394   & 1.742 & 0.654 & &       & 0.900 & 0.4635 & 0.0124 &  5384  & 1.745 & 0.659 \\
0.345 & 0.700 & 0.4663 & 0.0055 & 11081   & 1.789 & 0.660 & & 0.345 & 0.700 & 0.4663 & 0.0053 & 11069  & 1.794 & 0.665 \\
      & 0.800 & 0.4616 & 0.0092 &  6903   & 1.782 & 0.652 & &       & 0.800 & 0.4616 & 0.0091 &  6891  & 1.786 & 0.657 \\
      & 0.900 & 0.4564 & 0.0119 &  4597   & 1.770 & 0.643 & &       & 0.900 & 0.4562 & 0.0117 &  4587  & 1.773 & 0.648 \\
0.370 & 0.700 & 0.4599 & 0.0056 &  9361   & 1.824 & 0.651 & & 0.370 & 0.700 & 0.4599 & 0.0054 &  9349  & 1.828 & 0.656 \\
      & 0.800 & 0.4547 & 0.0088 &  5848   & 1.812 & 0.642 & &       & 0.800 & 0.4546 & 0.0087 &  5837  & 1.816 & 0.646 \\
      & 0.900 & 0.4488 & 0.0111 &  3906   & 1.797 & 0.631 & &       & 0.900 & 0.4484 & 0.0109 &  3896  & 1.799 & 0.635 \\
\midrule

\multicolumn{15}{c}{$Z = 0.00093$} \\
\midrule[\cmidrulewidth]
\multicolumn{7}{c}{$\FeH = -1.27, \aFe = 0.00$} & & \multicolumn{7}{c}{$\FeH = -1.50, \aFe = 0.30$} \\
\cmidrule{1-7} \cmidrule{9-15}
0.230 & 0.700 & 0.4911 & 0.0040 & 24044   & 1.609 & 0.654 & & 0.230 & 0.700 & 0.4913 & 0.0039 & 24037  & 1.614 & 0.659 \\
      & 0.800 & 0.4880 & 0.0104 & 14877   & 1.610 & 0.651 & &       & 0.800 & 0.4881 & 0.0101 & 14863  & 1.615 & 0.656 \\
      & 0.900 & 0.4849 & 0.0154 &  9780   & 1.610 & 0.647 & &       & 0.900 & 0.4850 & 0.0152 &  9761  & 1.614 & 0.653 \\
0.245 & 0.700 & 0.4877 & 0.0045 & 21942   & 1.628 & 0.651 & & 0.245 & 0.700 & 0.4878 & 0.0043 & 21935  & 1.633 & 0.656 \\
      & 0.800 & 0.4845 & 0.0106 & 13578   & 1.629 & 0.648 & &       & 0.800 & 0.4846 & 0.0103 & 13562  & 1.633 & 0.653 \\
      & 0.900 & 0.4814 & 0.0154 &  8931   & 1.628 & 0.644 & &       & 0.900 & 0.4814 & 0.0150 &  8914  & 1.632 & 0.649 \\
0.270 & 0.700 & 0.4819 & 0.0052 & 18789   & 1.661 & 0.646 & & 0.270 & 0.700 & 0.4821 & 0.0049 & 18779  & 1.666 & 0.651 \\
      & 0.800 & 0.4786 & 0.0108 & 11627   & 1.661 & 0.642 & &       & 0.800 & 0.4786 & 0.0105 & 11611  & 1.666 & 0.647 \\
      & 0.900 & 0.4753 & 0.0149 &  7662   & 1.658 & 0.638 & &       & 0.900 & 0.4753 & 0.0146 &  7645  & 1.662 & 0.643 \\
0.295 & 0.700 & 0.4761 & 0.0058 & 16029   & 1.695 & 0.640 & & 0.295 & 0.700 & 0.4762 & 0.0056 & 16016  & 1.700 & 0.644 \\
      & 0.800 & 0.4727 & 0.0108 &  9926   & 1.693 & 0.635 & &       & 0.800 & 0.4726 & 0.0106 &  9908  & 1.698 & 0.640 \\
      & 0.900 & 0.4692 & 0.0144 &  6556   & 1.689 & 0.630 & &       & 0.900 & 0.4690 & 0.0141 &  6538  & 1.693 & 0.635 \\
0.320 & 0.700 & 0.4703 & 0.0062 & 13623   & 1.730 & 0.633 & & 0.320 & 0.700 & 0.4703 & 0.0060 & 13608  & 1.735 & 0.637 \\
      & 0.800 & 0.4666 & 0.0107 &  8447   & 1.726 & 0.628 & &       & 0.800 & 0.4665 & 0.0104 &  8428  & 1.731 & 0.632 \\
      & 0.900 & 0.4628 & 0.0137 &  5594   & 1.720 & 0.622 & &       & 0.900 & 0.4626 & 0.0135 &  5576  & 1.724 & 0.626 \\
0.345 & 0.700 & 0.4644 & 0.0065 & 11535   & 1.765 & 0.624 & & 0.345 & 0.700 & 0.4644 & 0.0063 & 11518  & 1.770 & 0.629 \\
      & 0.800 & 0.4604 & 0.0103 &  7167   & 1.760 & 0.619 & &       & 0.800 & 0.4602 & 0.0101 &  7148  & 1.764 & 0.623 \\
      & 0.900 & 0.4561 & 0.0128 &  4760   & 1.750 & 0.612 & &       & 0.900 & 0.4558 & 0.0126 &  4743  & 1.754 & 0.616 \\
0.370 & 0.700 & 0.4584 & 0.0066 &  9731   & 1.801 & 0.616 & & 0.370 & 0.700 & 0.4583 & 0.0064 &  9711  & 1.806 & 0.620 \\
      & 0.800 & 0.4539 & 0.0098 &  6063   & 1.793 & 0.609 & &       & 0.800 & 0.4538 & 0.0096 &  6043  & 1.797 & 0.613 \\
      & 0.900 & 0.4493 & 0.0120 &  4039   & 1.781 & 0.602 & &       & 0.900 & 0.4492 & 0.0117 &  4022  & 1.785 & 0.606 \\
\midrule

\multicolumn{15}{c}{$Z = 0.00160$} \\
\midrule[\cmidrulewidth]
\multicolumn{7}{c}{$\FeH = -1.04, \aFe = 0.00$} & & \multicolumn{7}{c}{$\FeH = -1.26, \aFe = 0.30$} \\
\cmidrule{1-7} \cmidrule{9-15}
0.230 & 0.700 & 0.4886 & 0.0048 & 25702   & 1.578 & 0.627 & & 0.230 & 0.700 & 0.4887 & 0.0046 & 25695  & 1.583 & 0.632 \\
      & 0.800 & 0.4857 & 0.0114 & 15895   & 1.580 & 0.624 & &       & 0.800 & 0.4857 & 0.0112 & 15865  & 1.584 & 0.630 \\
      & 0.900 & 0.4829 & 0.0166 & 10425   & 1.581 & 0.622 & &       & 0.900 & 0.4829 & 0.0163 & 10389  & 1.585 & 0.626 \\
0.245 & 0.700 & 0.4853 & 0.0053 & 23443   & 1.598 & 0.624 & & 0.245 & 0.700 & 0.4854 & 0.0050 & 23428  & 1.603 & 0.629 \\
      & 0.800 & 0.4823 & 0.0117 & 14493   & 1.600 & 0.621 & &       & 0.800 & 0.4825 & 0.0114 & 14462  & 1.605 & 0.627 \\
      & 0.900 & 0.4795 & 0.0164 &  9510   & 1.599 & 0.618 & &       & 0.900 & 0.4795 & 0.0162 &  9473  & 1.604 & 0.623 \\
0.270 & 0.700 & 0.4798 & 0.0061 & 20051   & 1.632 & 0.619 & & 0.270 & 0.700 & 0.4798 & 0.0059 & 20028  & 1.638 & 0.623 \\
      & 0.800 & 0.4767 & 0.0119 & 12393   & 1.633 & 0.615 & &       & 0.800 & 0.4767 & 0.0116 & 12358  & 1.638 & 0.620 \\
      & 0.900 & 0.4738 & 0.0160 &  8141   & 1.631 & 0.612 & &       & 0.900 & 0.4737 & 0.0157 &  8105  & 1.636 & 0.617 \\
0.295 & 0.700 & 0.4742 & 0.0068 & 17086   & 1.668 & 0.613 & & 0.295 & 0.700 & 0.4742 & 0.0065 & 17057  & 1.673 & 0.617 \\
      & 0.800 & 0.4710 & 0.0119 & 10561   & 1.667 & 0.609 & &       & 0.800 & 0.4709 & 0.0116 & 10524  & 1.672 & 0.613 \\
      & 0.900 & 0.4680 & 0.0154 &  6950   & 1.664 & 0.605 & &       & 0.900 & 0.4678 & 0.0151 &  6915  & 1.668 & 0.610 \\
0.320 & 0.700 & 0.4686 & 0.0072 & 14504   & 1.705 & 0.606 & & 0.320 & 0.700 & 0.4686 & 0.0069 & 14469  & 1.711 & 0.610 \\
      & 0.800 & 0.4653 & 0.0117 &  8970   & 1.702 & 0.602 & &       & 0.800 & 0.4651 & 0.0114 &  8932  & 1.707 & 0.606 \\
      & 0.900 & 0.4620 & 0.0145 &  5918   & 1.697 & 0.598 & &       & 0.900 & 0.4618 & 0.0144 &  5884  & 1.702 & 0.602 \\
0.345 & 0.700 & 0.4630 & 0.0074 & 12263   & 1.742 & 0.599 & & 0.345 & 0.700 & 0.4629 & 0.0072 & 12225  & 1.748 & 0.603 \\
      & 0.800 & 0.4594 & 0.0112 &  7595   & 1.737 & 0.594 & &       & 0.800 & 0.4593 & 0.0110 &  7557  & 1.743 & 0.598 \\
      & 0.900 & 0.4560 & 0.0137 &  5025   & 1.731 & 0.590 & &       & 0.900 & 0.4557 & 0.0135 &  4993  & 1.735 & 0.593 \\
0.370 & 0.700 & 0.4573 & 0.0075 & 10328   & 1.781 & 0.591 & & 0.370 & 0.700 & 0.4572 & 0.0072 & 10287  & 1.786 & 0.595 \\
      & 0.800 & 0.4535 & 0.0107 &  6411   & 1.773 & 0.586 & &       & 0.800 & 0.4533 & 0.0105 &  6374  & 1.779 & 0.590 \\
      & 0.900 & 0.4498 & 0.0127 &  4255   & 1.765 & 0.581 & &       & 0.900 & 0.4494 & 0.0125 &  4225  & 1.769 & 0.584 \\
\midrule

\multicolumn{15}{c}{$Z = 0.00284$} \\
\midrule[\cmidrulewidth]
\multicolumn{7}{c}{$\FeH = -0.79, \aFe = 0.00$} & & \multicolumn{7}{c}{$\FeH = -1.01, \aFe = 0.30$} \\
\cmidrule{1-7} \cmidrule{9-15}
0.230 & 0.700 & 0.4862 & 0.0055 & 28644   & 1.540 & 0.603 & & 0.230 & 0.700 & 0.4863 & 0.0052 & 28618  & 1.544 & 0.607 \\
      & 0.800 & 0.4836 & 0.0125 & 17704   & 1.543 & 0.601 & &       & 0.800 & 0.4836 & 0.0122 & 17652  & 1.547 & 0.606 \\
      & 0.900 & 0.4812 & 0.0176 & 11584   & 1.544 & 0.599 & &       & 0.900 & 0.4811 & 0.0174 & 11523  & 1.549 & 0.604 \\
0.245 & 0.700 & 0.4831 & 0.0060 & 26099   & 1.560 & 0.601 & & 0.245 & 0.700 & 0.4831 & 0.0058 & 26061  & 1.565 & 0.605 \\
      & 0.800 & 0.4804 & 0.0128 & 16126   & 1.563 & 0.598 & &       & 0.800 & 0.4803 & 0.0124 & 16068  & 1.568 & 0.603 \\
      & 0.900 & 0.4778 & 0.0175 & 10549   & 1.563 & 0.595 & &       & 0.900 & 0.4779 & 0.0173 & 10488  & 1.569 & 0.601 \\
0.270 & 0.700 & 0.4778 & 0.0069 & 22289   & 1.596 & 0.595 & & 0.270 & 0.700 & 0.4778 & 0.0066 & 22234  & 1.602 & 0.600 \\
      & 0.800 & 0.4750 & 0.0129 & 13761   & 1.598 & 0.593 & &       & 0.800 & 0.4750 & 0.0127 & 13696  & 1.603 & 0.597 \\
      & 0.900 & 0.4726 & 0.0170 &  9004   & 1.598 & 0.591 & &       & 0.900 & 0.4724 & 0.0167 &  8941  & 1.602 & 0.595 \\
0.295 & 0.700 & 0.4724 & 0.0075 & 18962   & 1.634 & 0.590 & & 0.295 & 0.700 & 0.4724 & 0.0074 & 18895  & 1.640 & 0.594 \\
      & 0.800 & 0.4696 & 0.0128 & 11699   & 1.633 & 0.587 & &       & 0.800 & 0.4695 & 0.0126 & 11631  & 1.640 & 0.591 \\
      & 0.900 & 0.4671 & 0.0163 &  7662   & 1.632 & 0.584 & &       & 0.900 & 0.4669 & 0.0161 &  7600  & 1.637 & 0.588 \\
0.320 & 0.700 & 0.4671 & 0.0081 & 16068   & 1.673 & 0.584 & & 0.320 & 0.700 & 0.4670 & 0.0078 & 15995  & 1.679 & 0.587 \\
      & 0.800 & 0.4642 & 0.0125 &  9911   & 1.671 & 0.581 & &       & 0.800 & 0.4641 & 0.0123 &  9841  & 1.677 & 0.584 \\
      & 0.900 & 0.4617 & 0.0155 &  6502   & 1.667 & 0.577 & &       & 0.900 & 0.4613 & 0.0153 &  6443  & 1.673 & 0.581 \\
0.345 & 0.700 & 0.4618 & 0.0083 & 13559   & 1.712 & 0.577 & & 0.345 & 0.700 & 0.4616 & 0.0081 & 13482  & 1.719 & 0.580 \\
      & 0.800 & 0.4588 & 0.0121 &  8366   & 1.709 & 0.574 & &       & 0.800 & 0.4586 & 0.0120 &  8298  & 1.715 & 0.577 \\
      & 0.900 & 0.4560 & 0.0144 &  5503   & 1.704 & 0.570 & &       & 0.900 & 0.4556 & 0.0143 &  5447  & 1.709 & 0.574 \\
0.370 & 0.700 & 0.4564 & 0.0084 & 11393   & 1.753 & 0.570 & & 0.370 & 0.700 & 0.4562 & 0.0082 & 11315  & 1.760 & 0.573 \\
      & 0.800 & 0.4532 & 0.0115 &  7039   & 1.747 & 0.566 & &       & 0.800 & 0.4530 & 0.0114 &  6973  & 1.754 & 0.570 \\
      & 0.900 & 0.4503 & 0.0133 &  4644   & 1.743 & 0.563 & &       & 0.900 & 0.4499 & 0.0133 &  4592  & 1.747 & 0.565 \\
\midrule

\multicolumn{15}{c}{$Z = 0.00503$} \\
\midrule[\cmidrulewidth]
\multicolumn{7}{c}{$\FeH = -0.54, \aFe = 0.00$} & & \multicolumn{7}{c}{$\FeH = -0.76, \aFe = 0.30$} \\
\cmidrule{1-7} \cmidrule{9-15}
0.230 & 0.700 & 0.4839 & 0.0059 & 33629   & 1.493 & 0.583 & & 0.230 & 0.700 & 0.4839 & 0.0057 & 33567  & 1.497 & 0.587 \\
      & 0.800 & 0.4814 & 0.0132 & 20761   & 1.497 & 0.581 & &       & 0.800 & 0.4814 & 0.0130 & 20669  & 1.501 & 0.585 \\
      & 0.900 & 0.4793 & 0.0185 & 13556   & 1.498 & 0.579 & &       & 0.900 & 0.4792 & 0.0183 & 13458  & 1.503 & 0.584 \\
0.245 & 0.700 & 0.4807 & 0.0070 & 30609   & 1.514 & 0.580 & & 0.245 & 0.700 & 0.4808 & 0.0063 & 30518  & 1.518 & 0.585 \\
      & 0.800 & 0.4784 & 0.0134 & 18886   & 1.516 & 0.578 & &       & 0.800 & 0.4783 & 0.0133 & 18786  & 1.522 & 0.583 \\
      & 0.900 & 0.4762 & 0.0184 & 12324   & 1.517 & 0.577 & &       & 0.900 & 0.4761 & 0.0181 & 12223  & 1.524 & 0.581 \\
0.270 & 0.700 & 0.4757 & 0.0075 & 26067   & 1.550 & 0.576 & & 0.270 & 0.700 & 0.4757 & 0.0073 & 25965  & 1.557 & 0.580 \\
      & 0.800 & 0.4733 & 0.0138 & 16081   & 1.554 & 0.574 & &       & 0.800 & 0.4732 & 0.0135 & 15970  & 1.559 & 0.578 \\
      & 0.900 & 0.4712 & 0.0179 & 10484   & 1.554 & 0.572 & &       & 0.900 & 0.4710 & 0.0176 & 10380  & 1.558 & 0.576 \\
0.295 & 0.700 & 0.4707 & 0.0083 & 22128   & 1.591 & 0.571 & & 0.295 & 0.700 & 0.4706 & 0.0081 & 22011  & 1.597 & 0.575 \\
      & 0.800 & 0.4682 & 0.0137 & 13637   & 1.591 & 0.568 & &       & 0.800 & 0.4681 & 0.0135 & 13521  & 1.599 & 0.573 \\
      & 0.900 & 0.4660 & 0.0171 &  8888   & 1.590 & 0.566 & &       & 0.900 & 0.4658 & 0.0169 &  8785  & 1.597 & 0.570 \\
0.320 & 0.700 & 0.4656 & 0.0089 & 18706   & 1.632 & 0.565 & & 0.320 & 0.700 & 0.4655 & 0.0087 & 18583  & 1.639 & 0.568 \\
      & 0.800 & 0.4631 & 0.0135 & 11517   & 1.632 & 0.563 & &       & 0.800 & 0.4629 & 0.0133 & 11401  & 1.638 & 0.566 \\
      & 0.900 & 0.4609 & 0.0163 &  7510   & 1.629 & 0.560 & &       & 0.900 & 0.4606 & 0.0161 &  7412  & 1.635 & 0.564 \\
0.345 & 0.700 & 0.4605 & 0.0092 & 15747   & 1.673 & 0.559 & & 0.345 & 0.700 & 0.4604 & 0.0090 & 15619  & 1.682 & 0.562 \\
      & 0.800 & 0.4580 & 0.0130 &  9687   & 1.671 & 0.556 & &       & 0.800 & 0.4577 & 0.0129 &  9575  & 1.679 & 0.560 \\
      & 0.900 & 0.4558 & 0.0153 &  6327   & 1.668 & 0.554 & &       & 0.900 & 0.4554 & 0.0151 &  6234  & 1.674 & 0.557 \\
0.370 & 0.700 & 0.4555 & 0.0092 & 13197   & 1.717 & 0.553 & & 0.370 & 0.700 & 0.4553 & 0.0090 & 13070  & 1.725 & 0.555 \\
      & 0.800 & 0.4528 & 0.0124 &  8117   & 1.714 & 0.550 & &       & 0.800 & 0.4526 & 0.0122 &  8010  & 1.722 & 0.553 \\
      & 0.900 & 0.4506 & 0.0141 &  5315   & 1.709 & 0.547 & &       & 0.900 & 0.4502 & 0.0140 &  5230  & 1.717 & 0.550 \\
\midrule

\multicolumn{15}{c}{$Z = 0.00890$} \\
\midrule[\cmidrulewidth]
\multicolumn{7}{c}{$\FeH = -0.29, \aFe = 0.00$} & & \multicolumn{7}{c}{$\FeH = -0.51, \aFe = 0.30$} \\
\cmidrule{1-7} \cmidrule{9-15}
0.230 & 0.700 & 0.4810 & 0.0061 & 41905   & 1.433 & 0.564 & & 0.230 & 0.700 & 0.4812 & 0.0058 & 41781  & 1.439 & 0.569 \\
      & 0.800 & 0.4789 & 0.0138 & 25811   & 1.439 & 0.563 & &       & 0.800 & 0.4788 & 0.0135 & 25647  & 1.443 & 0.567 \\
      & 0.900 & 0.4770 & 0.0192 & 16827   & 1.442 & 0.561 & &       & 0.900 & 0.4769 & 0.0190 & 16662  & 1.446 & 0.566 \\
0.245 & 0.700 & 0.4783 & 0.0069 & 38038   & 1.455 & 0.563 & & 0.245 & 0.700 & 0.4783 & 0.0066 & 37899  & 1.459 & 0.567 \\
      & 0.800 & 0.4760 & 0.0142 & 23433   & 1.459 & 0.561 & &       & 0.800 & 0.4760 & 0.0140 & 23264  & 1.465 & 0.565 \\
      & 0.900 & 0.4741 & 0.0190 & 15269   & 1.461 & 0.559 & &       & 0.900 & 0.4740 & 0.0190 & 15102  & 1.468 & 0.564 \\
0.270 & 0.700 & 0.4734 & 0.0079 & 32285   & 1.492 & 0.559 & & 0.270 & 0.700 & 0.4734 & 0.0078 & 32124  & 1.500 & 0.563 \\
      & 0.800 & 0.4712 & 0.0145 & 19889   & 1.496 & 0.556 & &       & 0.800 & 0.4711 & 0.0143 & 19712  & 1.504 & 0.561 \\
      & 0.900 & 0.4693 & 0.0187 & 12945   & 1.497 & 0.555 & &       & 0.900 & 0.4691 & 0.0185 & 12777  & 1.504 & 0.560 \\
0.295 & 0.700 & 0.4686 & 0.0088 & 27307   & 1.534 & 0.554 & & 0.295 & 0.700 & 0.4685 & 0.0086 & 27127  & 1.541 & 0.558 \\
      & 0.800 & 0.4664 & 0.0145 & 16818   & 1.537 & 0.552 & &       & 0.800 & 0.4662 & 0.0144 & 16636  & 1.543 & 0.556 \\
      & 0.900 & 0.4645 & 0.0179 & 10930   & 1.537 & 0.551 & &       & 0.900 & 0.4643 & 0.0178 & 10767  & 1.544 & 0.554 \\
0.320 & 0.700 & 0.4637 & 0.0095 & 23014   & 1.578 & 0.549 & & 0.320 & 0.700 & 0.4636 & 0.0093 & 22821  & 1.587 & 0.553 \\
      & 0.800 & 0.4615 & 0.0142 & 14160   & 1.579 & 0.547 & &       & 0.800 & 0.4614 & 0.0141 & 13979  & 1.588 & 0.551 \\
      & 0.900 & 0.4597 & 0.0170 &  9193   & 1.577 & 0.545 & &       & 0.900 & 0.4594 & 0.0169 &  9035  & 1.585 & 0.549 \\
0.345 & 0.700 & 0.4589 & 0.0099 & 19318   & 1.623 & 0.543 & & 0.345 & 0.700 & 0.4588 & 0.0096 & 19118  & 1.633 & 0.547 \\
      & 0.800 & 0.4567 & 0.0138 & 11872   & 1.622 & 0.541 & &       & 0.800 & 0.4565 & 0.0136 & 11693  & 1.631 & 0.545 \\
      & 0.900 & 0.4548 & 0.0159 &  7704   & 1.619 & 0.539 & &       & 0.900 & 0.4545 & 0.0159 &  7555  & 1.628 & 0.543 \\
0.370 & 0.700 & 0.4542 & 0.0100 & 16145   & 1.677 & 0.537 & & 0.370 & 0.700 & 0.4540 & 0.0098 & 15944  & 1.680 & 0.541 \\
      & 0.800 & 0.4519 & 0.0132 &  9908   & 1.668 & 0.535 & &       & 0.800 & 0.4517 & 0.0131 &  9735  & 1.678 & 0.538 \\
      & 0.900 & 0.4500 & 0.0147 &  6436   & 1.664 & 0.533 & &       & 0.900 & 0.4497 & 0.0148 &  6296  & 1.674 & 0.536 \\
\midrule

\multicolumn{15}{c}{$Z = 0.01570$} \\
\midrule[\cmidrulewidth]
\multicolumn{7}{c}{$\FeH = -0.04, \aFe = 0.00$} & & \multicolumn{7}{c}{$\FeH = -0.26, \aFe = 0.30$} \\
\cmidrule{1-7} \cmidrule{9-15}
0.230 & 0.700 & 0.4776 & 0.0060 & 55021   & 1.366 & 0.546 & & 0.230 & 0.700 & 0.4776 & 0.0059 & 54953  & 1.373 & 0.551 \\
      & 0.800 & 0.4754 & 0.0142 & 33842   & 1.371 & 0.545 & &       & 0.800 & 0.4755 & 0.0140 & 33651  & 1.378 & 0.550 \\
      & 0.900 & 0.4736 & 0.0201 & 22014   & 1.374 & 0.544 & &       & 0.900 & 0.4736 & 0.0199 & 21790  & 1.381 & 0.549 \\
0.245 & 0.700 & 0.4748 & 0.0069 & 49854   & 1.385 & 0.545 & & 0.245 & 0.700 & 0.4748 & 0.0067 & 49739  & 1.391 & 0.550 \\
      & 0.800 & 0.4727 & 0.0147 & 30643   & 1.390 & 0.544 & &       & 0.800 & 0.4727 & 0.0144 & 30437  & 1.396 & 0.548 \\
      & 0.900 & 0.4709 & 0.0199 & 19933   & 1.393 & 0.543 & &       & 0.900 & 0.4709 & 0.0199 & 19702  & 1.400 & 0.547 \\
0.270 & 0.700 & 0.4701 & 0.0082 & 42141   & 1.421 & 0.542 & & 0.270 & 0.700 & 0.4702 & 0.0080 & 41969  & 1.429 & 0.547 \\
      & 0.800 & 0.4681 & 0.0151 & 25900   & 1.428 & 0.541 & &       & 0.800 & 0.4680 & 0.0150 & 25670  & 1.436 & 0.545 \\
      & 0.900 & 0.4663 & 0.0195 & 16840   & 1.429 & 0.539 & &       & 0.900 & 0.4663 & 0.0195 & 16604  & 1.438 & 0.544 \\
0.295 & 0.700 & 0.4655 & 0.0092 & 35491   & 1.464 & 0.538 & & 0.295 & 0.700 & 0.4655 & 0.0091 & 35273  & 1.473 & 0.543 \\
      & 0.800 & 0.4635 & 0.0153 & 21812   & 1.468 & 0.537 & &       & 0.800 & 0.4634 & 0.0152 & 21570  & 1.477 & 0.541 \\
      & 0.900 & 0.4618 & 0.0188 & 14169   & 1.469 & 0.535 & &       & 0.900 & 0.4617 & 0.0188 & 13933  & 1.478 & 0.540 \\
0.320 & 0.700 & 0.4609 & 0.0100 & 29775   & 1.509 & 0.534 & & 0.320 & 0.700 & 0.4609 & 0.0099 & 29529  & 1.520 & 0.538 \\
      & 0.800 & 0.4589 & 0.0150 & 18302   & 1.514 & 0.533 & &       & 0.800 & 0.4588 & 0.0150 & 18049  & 1.522 & 0.536 \\
      & 0.900 & 0.4574 & 0.0178 & 11872   & 1.512 & 0.531 & &       & 0.900 & 0.4571 & 0.0179 & 11640  & 1.522 & 0.535 \\
0.345 & 0.700 & 0.4564 & 0.0105 & 24886   & 1.558 & 0.529 & & 0.345 & 0.700 & 0.4563 & 0.0104 & 24621  & 1.570 & 0.532 \\
      & 0.800 & 0.4545 & 0.0146 & 15292   & 1.558 & 0.527 & &       & 0.800 & 0.4542 & 0.0146 & 15036  & 1.570 & 0.531 \\
      & 0.900 & 0.4528 & 0.0167 &  9904   & 1.555 & 0.525 & &       & 0.900 & 0.4526 & 0.0168 &  9679  & 1.568 & 0.529 \\
0.370 & 0.700 & 0.4519 & 0.0107 & 20718   & 1.608 & 0.524 & & 0.370 & 0.700 & 0.4517 & 0.0106 & 20440  & 1.619 & 0.527 \\
      & 0.800 & 0.4499 & 0.0140 & 12718   & 1.607 & 0.522 & &       & 0.800 & 0.4496 & 0.0139 & 12466  & 1.620 & 0.525 \\
      & 0.900 & 0.4482 & 0.0155 &  8227   & 1.607 & 0.520 & &       & 0.900 & 0.4480 & 0.0156 &  8014  & 1.620 & 0.523 \\
\midrule

\multicolumn{15}{c}{$Z = 0.03000$} \\
\midrule[\cmidrulewidth]
\multicolumn{7}{c}{$\FeH = 0.25, \aFe = 0.00$} & & \multicolumn{7}{c}{$\FeH = 0.03, \aFe = 0.30$} \\
\cmidrule{1-7} \cmidrule{9-15}
0.230 & 0.700 & 0.4711 & 0.0064 & 77421   & 1.279 & 0.526 & & 0.230 & 0.700 & 0.4712 & 0.0062 & 77230  & 1.295 & 0.529 \\
      & 0.800 & 0.4691 & 0.0148 & 48185   & 1.283 & 0.525 & &       & 0.800 & 0.4692 & 0.0147 & 47811  & 1.298 & 0.528 \\
      & 0.900 & 0.4674 & 0.0212 & 31263   & 1.285 & 0.524 & &       & 0.900 & 0.4675 & 0.0212 & 30872  & 1.300 & 0.527 \\
0.245 & 0.700 & 0.4686 & 0.0073 & 70264   & 1.295 & 0.525 & & 0.245 & 0.700 & 0.4685 & 0.0072 & 69990  & 1.309 & 0.528 \\
      & 0.800 & 0.4664 & 0.0154 & 43501   & 1.299 & 0.524 & &       & 0.800 & 0.4665 & 0.0153 & 43103  & 1.313 & 0.527 \\
      & 0.900 & 0.4649 & 0.0210 & 28201   & 1.301 & 0.523 & &       & 0.900 & 0.4649 & 0.0211 & 27807  & 1.317 & 0.526 \\
0.270 & 0.700 & 0.4641 & 0.0086 & 59395   & 1.327 & 0.523 & & 0.270 & 0.700 & 0.4641 & 0.0086 & 59027  & 1.343 & 0.526 \\
      & 0.800 & 0.4621 & 0.0161 & 36550   & 1.333 & 0.522 & &       & 0.800 & 0.4621 & 0.0161 & 36130  & 1.349 & 0.525 \\
      & 0.900 & 0.4606 & 0.0206 & 23680   & 1.336 & 0.521 & &       & 0.900 & 0.4606 & 0.0207 & 23286  & 1.352 & 0.524 \\
0.295 & 0.700 & 0.4597 & 0.0099 & 49854   & 1.370 & 0.520 & & 0.295 & 0.700 & 0.4597 & 0.0099 & 49430  & 1.386 & 0.523 \\
      & 0.800 & 0.4578 & 0.0164 & 30585   & 1.375 & 0.519 & &       & 0.800 & 0.4578 & 0.0164 & 30159  & 1.392 & 0.522 \\
      & 0.900 & 0.4563 & 0.0200 & 19809   & 1.376 & 0.517 & &       & 0.900 & 0.4563 & 0.0200 & 19422  & 1.394 & 0.521 \\
0.320 & 0.700 & 0.4553 & 0.0109 & 41592   & 1.416 & 0.516 & & 0.320 & 0.700 & 0.4553 & 0.0108 & 41142  & 1.434 & 0.519 \\
      & 0.800 & 0.4535 & 0.0163 & 25491   & 1.421 & 0.515 & &       & 0.800 & 0.4535 & 0.0162 & 25066  & 1.438 & 0.518 \\
      & 0.900 & 0.4521 & 0.0189 & 16503   & 1.421 & 0.513 & &       & 0.900 & 0.4521 & 0.0191 & 16130  & 1.439 & 0.517 \\
0.345 & 0.700 & 0.4510 & 0.0115 & 34534   & 1.468 & 0.512 & & 0.345 & 0.700 & 0.4510 & 0.0115 & 34066  & 1.488 & 0.515 \\
      & 0.800 & 0.4493 & 0.0158 & 21162   & 1.470 & 0.511 & &       & 0.800 & 0.4493 & 0.0157 & 20746  & 1.490 & 0.513 \\
      & 0.900 & 0.4479 & 0.0178 & 13694   & 1.470 & 0.509 & &       & 0.900 & 0.4479 & 0.0179 & 13335  & 1.489 & 0.512 \\
0.370 & 0.700 & 0.4467 & 0.0117 & 28544   & 1.522 & 0.507 & & 0.370 & 0.700 & 0.4467 & 0.0117 & 28078  & 1.544 & 0.510 \\
      & 0.800 & 0.4450 & 0.0150 & 17495   & 1.523 & 0.506 & &       & 0.800 & 0.4450 & 0.0151 & 17093  & 1.542 & 0.508 \\
      & 0.900 & 0.4437 & 0.0164 & 11311   & 1.522 & 0.504 & &       & 0.900 & 0.4436 & 0.0165 & 10971  & 1.543 & 0.507 \\
\bottomrule
\end{longtable}
}

\let\clearpage\relax
\vspace*{24pt}
\section{Supplementary HB track figures}
\label{sec:extrafigures}

\begin{figure*}
    \centering
        \includegraphics[width=1.95\columnwidth]{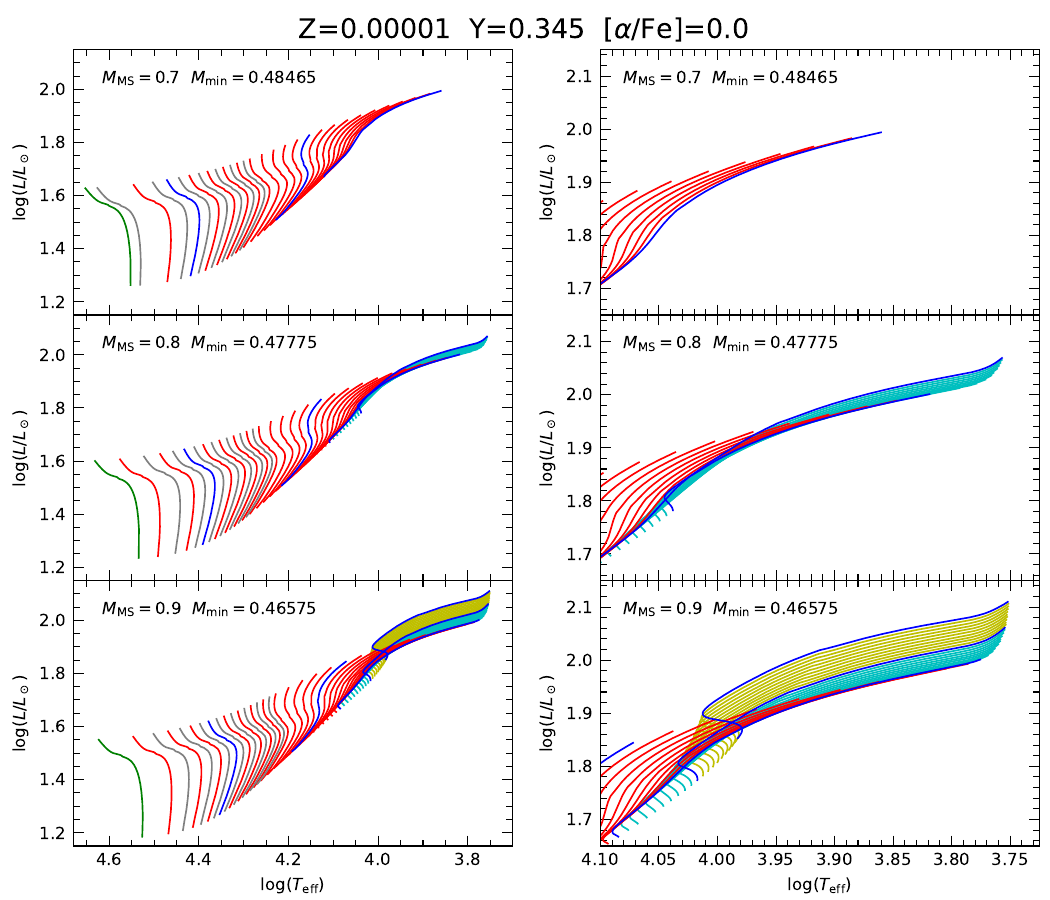}    
    \caption{Same as Fig.~\ref{Fig:HB1} but for $Y=0.345$.}
    \label{Fig:HB2}
\end{figure*}

\begin{figure*}
    \centering
        \includegraphics[width=1.95\columnwidth]{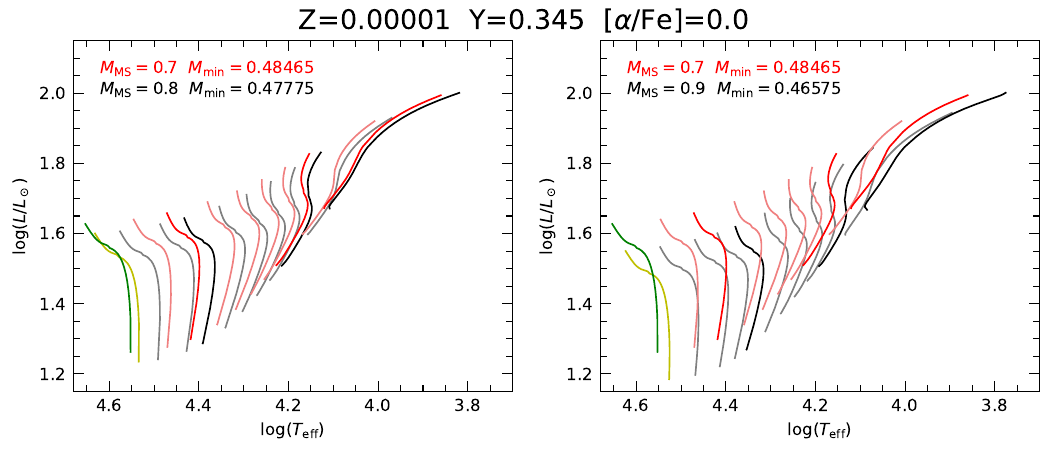}
    \caption{Same as Fig.~\ref{Fig:HB1b} but for $Y=0.345$.}
    \label{Fig:HB2b}
\end{figure*}

\begin{figure*}
    \centering
    \includegraphics[width=1.95\columnwidth]{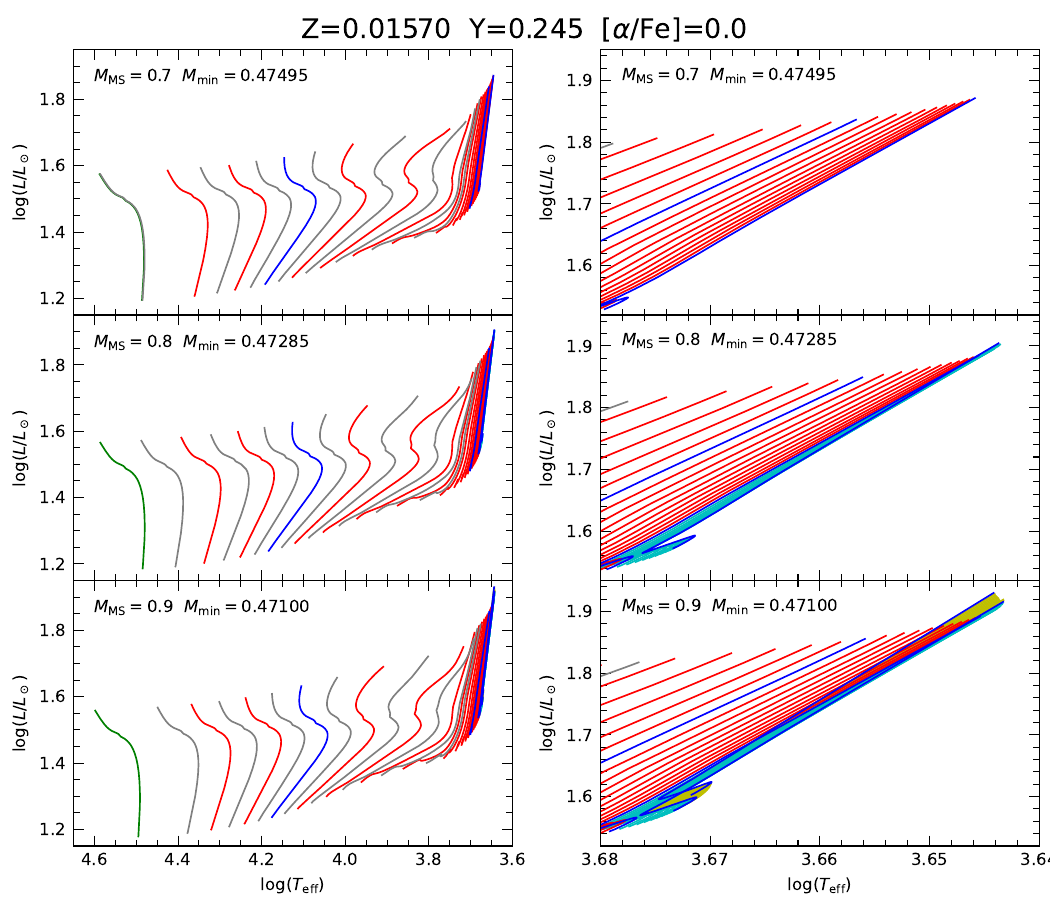}
    \caption{Same as Fig.~\ref{Fig:HB1} but for $Z=0.01570$ and $Y=0.245$.}
    \label{Fig:HB3}
\end{figure*}

\begin{figure*}
    \centering
        \includegraphics[width=1.95\columnwidth]{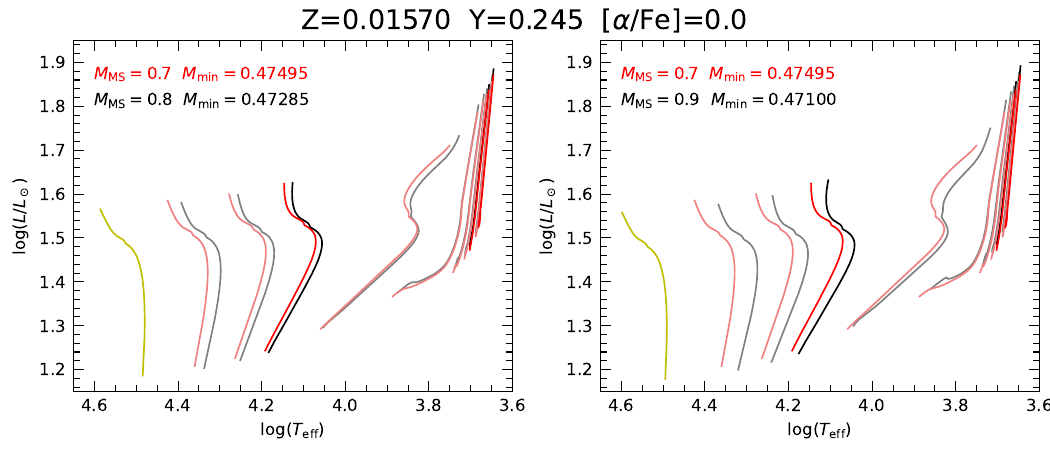}    
    \caption{Same as Fig.~\ref{Fig:HB1b} but for $Z=0.01570$ and $Y=0.245$.}
    \label{Fig:HB3b}
\end{figure*}

\begin{figure*}
    \centering
        \includegraphics[width=1.95\columnwidth]{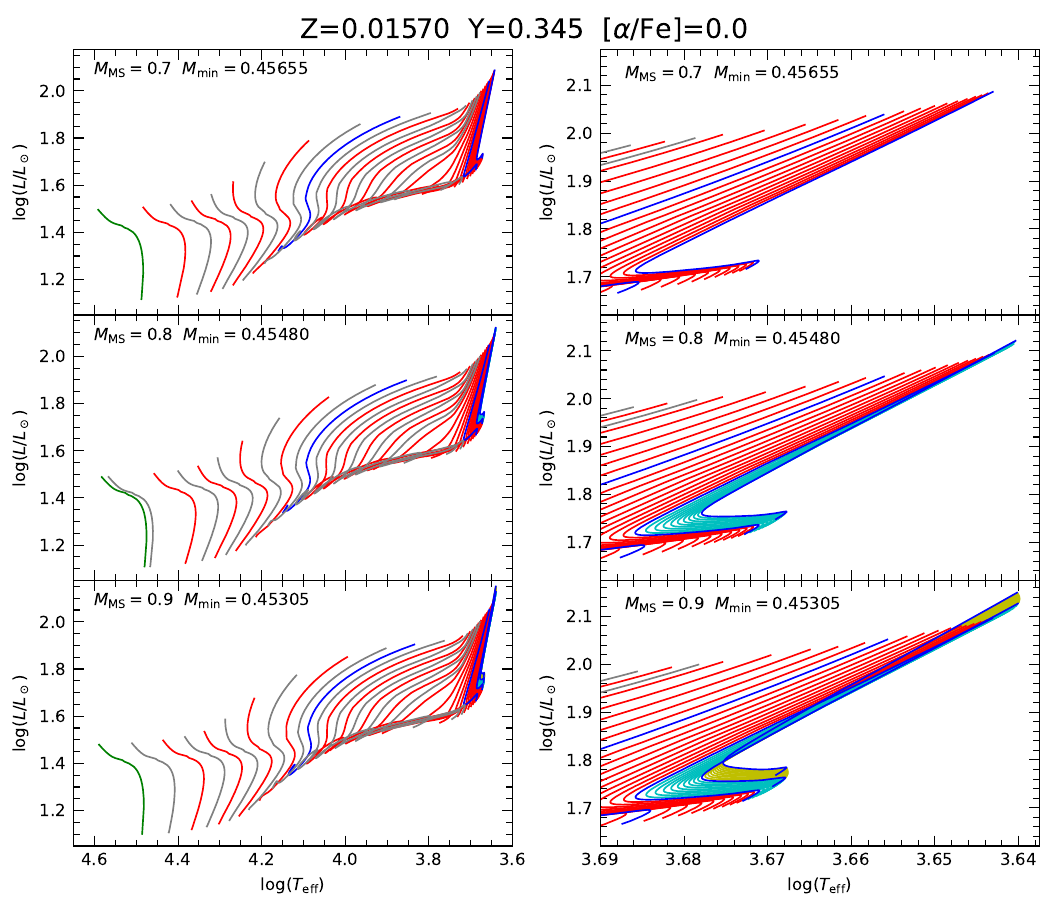}
    \caption{Same as Fig.~\ref{Fig:HB1} but for $Z=0.01570$ and $Y=0.345$.}
    \label{Fig:HB4}
\end{figure*}

\begin{figure*}
    \centering
        \includegraphics[width=1.95\columnwidth]{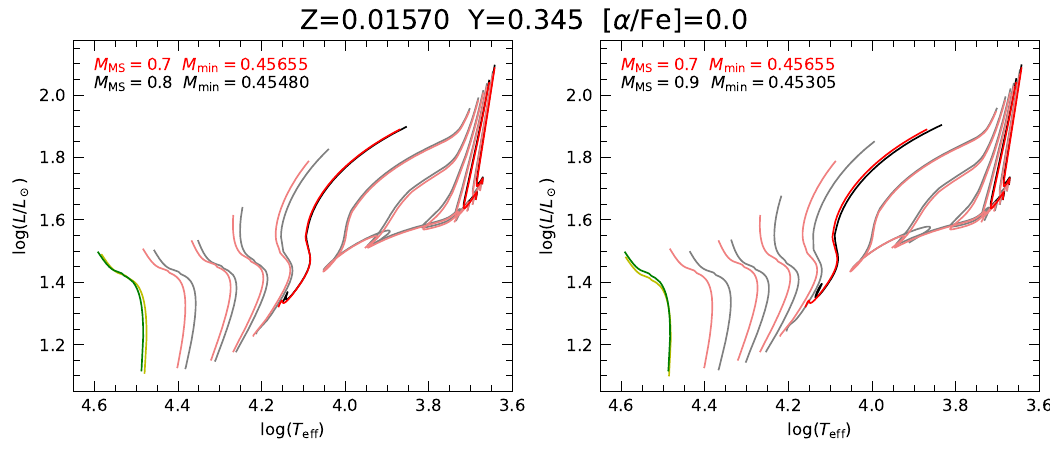} 
    \caption{Same as Fig.~\ref{Fig:HB1b} but for $Z=0.01570$ and $Y=0.345$.}
    \label{Fig:HB4b}
\end{figure*}

To complement the example presented in Figs.~\ref{Fig:HB1} and ~\ref{Fig:HB1b}, the figures in this appendix display three additional sets of computed HB tracks for different chemical compositions and the three $\MMS$ values. Each set includes, similar to Fig.~\ref{Fig:HB1b}, a comparison between HB tracks with $\MMS=0.800$ and $0.900 \, \Mo$ against $0.700 \, \Mo$ for the corresponding chemical composition. Specifically, Figs.~\ref{Fig:HB2} and \ref{Fig:HB2b} correspond to the $Z=0.00001$, $Y=0.345$, and $\aFe=0.0$ case; Figs.~\ref{Fig:HB3} and \ref{Fig:HB3b}, to the $Z=0.01570$, $Y=0.245$, and $\aFe=0.0$ case; and Figs.~\ref{Fig:HB4} and \ref{Fig:HB4b}, to the $Z=0.01570$, $Y=0.345$, and $\aFe=0.0$ combination. 

\twocolumn




\end{appendix}

\end{document}